\documentclass[%
twocolumn,
showpacs,
preprintnumbers,
aps,
prd,
a4paper,
superscriptaddress,
nofootinbib,
amsmath,amssymb
floats,floatfix
]{revtex4-1}

\usepackage{parskip}
\usepackage[usenames,dvipsnames,svgnames,table]{xcolor}
\usepackage[colorlinks=true,linkcolor=MidnightBlue,citecolor=RawSienna,urlcolor=MidnightBlue]{hyperref}
\usepackage{graphicx}
\usepackage{natbib}
\usepackage{lmodern}

\graphicspath{ {Figures/} }

\usepackage{amsfonts}
\usepackage{amssymb}
\usepackage{amsmath}

\usepackage{mathrsfs}

\usepackage{bm}

\usepackage{color}
\usepackage[labelfont=sf,font+=small,format=hang,justification=RaggedRight]{subcaption}
\renewcommand{\d}{\mathrm{d}}

\renewcommand{\leq}{\leqslant}
\newcommand{\e}[1]{\mathrm{e}^{{#1}}}

\newcommand{\vect}[1]{\bm{\mathrm{{#1}}}}
\newcommand{\grad}{\nabla}

\newcommand{\Mp}{M_{\mathrm{P}}}

\newcommand{\kpiv}{k_{0}}

\newcommand{\Lik}{\mathscr{L}}

\newcommand{\feff}{f_{\text{eff}}}
\newcommand{\feffNewton}{f_{\text{eff}}^\Phi}
\newcommand{\feffGalileon}{f_{\text{eff}}^\phi}

\newcommand{\deltaobs}{\delta_{\obs}}
\newcommand{\obs}{\text{obs}}

\newcommand{\zeq}{z_{\text{eq}}}
\newcommand{\zdrag}{z_{\text{drag}}}
\newcommand{\fbaryon}{f_{\text{baryon}}}

\newcommand{\Lbreak}{\mathcal{L}_{\text{break}}}
\newcommand{\Lag}{\mathcal{L}}

\newcommand{\GNewton}{G}
\newcommand{\GeV}{\text{GeV}}
\newcommand{\eV}{\text{eV}}
\newcommand{\Mpc}{\text{Mpc}}

\newcommand{\zeff}{z_{\text{eff}}}

\newcommand{\Pcon}{P^{\text{con}}}
\newcommand{\Pmodel}{\bm{\mathrm{P}}^{\text{mod}}}

\newcommand{\Pdata}{\bm{\mathrm{P}}^{\text{data}}}

\newcommand{\Cov}{\bm{\mathrm{C}}}

\newcommand{\discrepancy}{\bm{\mathrm{\delta}}}

\newcommand{\LCDM}{\Lambda\text{CDM}}

\newcommand{\Msource}{M_{\mathrm{c}}}
\newcommand{\Rsource}{R_{\mathrm{c}}}
\newcommand{\rhosource}{\rho_{\mathrm{c}}}
\newcommand{\RVainshtein}{R_{\mathrm{v}}}

\newcommand{\Gammaeff}{\Gamma_{\text{eff}}}

\newcommand{\Legendre}[1]{\mathscr{P}_{{#1}}}

\newcommand{\transfer}[2]{{T_{{#2}}}^{{#1}}}
\newcommand{\Coupling}{M}
\newcommand{\SelfInteraction}{\Lambda}

\newcommand{\AmplitudeNewton}{A_\Phi}
\newcommand{\AmplitudeGalileon}{A_\phi}

\newcommand{\transpose}{\mathsf{T}}

\newcommand{\semibold}[1]{{\fontseries{b}\selectfont{#1}}}

\newcommand{\para}[1]{\par\vspace{2mm}\noindent\semibold{{#1.}---}\ignorespaces}

\newcommand{\sidehead}[1]{\par\vspace{1.5mm}\noindent\textsf{{#1}.---}\ignorespaces}

\begin{document}

\title{Beyond the growth rate of cosmic structure: \\
Testing modified gravity models with an extra degree of freedom}
\author{Clare~Burrage}
\email{Clare.Burrage@nottingham.ac.uk}
\affiliation{School of Physics and Astronomy, University of Nottingham,
Nottingham, NG7 2RD, United Kingdom}
\author{David~Parkinson}
\email{d.parkinson@uq.edu.au}
\affiliation{School of Mathematics and Physics, University of Queensland, Brisbane, QLD 4072, Australia}
\author{David~Seery}
\email{D.Seery@sussex.ac.uk}
\affiliation{Department of Physics and Astronomy, University of Sussex, Brighton BN1 9QH, United Kingdom}

\begin{abstract}
In `modified' gravity the observed
acceleration of the universe is explained
by changing the gravitational force law
or the number of degrees of freedom in the gravitational sector.
Both possibilities can be tested by measurements of
cosmological structure formation.
In this paper we elaborate the details of such tests
using the Galileon model as a case study.
We pay attention to the possibility that each new degree of
freedom may have stochastically independent initial conditions,
generating different types of potential well in the early universe and
breaking complete correlation between density and velocity
power spectra. This `stochastic bias'
can confuse
schemes to parametrize the predictions of modified
gravity models,
such as the use of the growth parameter $f$ alone.
Using
data from the WiggleZ Dark Energy Survey
we show that it will be possible to obtain constraints
using information about the cosmological-scale force
law embedded in the multipole power spectra
of redshift-space distortions.
As an example,
we obtain an upper limit
on the strength of the conformal coupling to
matter in the cubic Galileon model,
giving
$|1/M| \lesssim 200 / \Mp$.
This allows the fifth-force to be
stronger than gravity, but is
consistent with zero coupling.
\end{abstract}

\maketitle

\section{Introduction}
The measurement of accelerated expansion
is one of
the most important discoveries in modern cosmology.
But although
this conclusion is
now supported by multiple
probes~\cite{jla2014,WiggleZBAO,BOSSDR11BAO,BOSSLyaBAO,Hinshaw:2012aka,Ade:2013zuv},
we remain entirely ignorant of the underlying physics.

A microphysical explanation could arise
from quantum fluctuations of the vacuum,
which
behave as a cosmological constant---a
fluid with equation of state $p = -\rho$.
But if this is the correct microphysical description
it will be difficult to understand why the universe
is accelerating so slowly.
This has encouraged the growth of a large literature
studying alternative explanations (for a review of such
ideas, see Ref.~\cite{Bull:2015stt}).

If vacuum fluctuations are rejected,
it seems inevitable that the acceleration must
depend on new physics operating over
very large scales which is not present
in Einstein gravity.
It is much less clear what this new physics
should be,
but at sufficiently low energies
it is likely to appear to us in the form
of extra scalar fields.
These mediate
long-range forces which compete with, or augment,
the conventional $1/r^2$ gravitational force.

Any modification to the
force law would influence the
assembly of cosmic structures.
Therefore, even if spatially-averaged properties---such
as the time history of the expansion rate $H$---%
can be made to agree with a $\LCDM$ model,
the details of structure formation will often disagree.
This makes measurements of structure formation a
discriminating test of modified gravities.
A sizeable industry has emerged
which aims to detect
departures from Einstein gravity
by
using surveys of the
cosmological density and velocity fields
to constrain the force law.

In modifications of gravity which entail extra fields
there is another effect
which is less frequently considered.
According to current ideas,
structure on all scales was seeded by quantum fluctuations
which became imprinted on the gravitational potential $\Phi$
during an early inflationary epoch.
The same effect will occur for every field which is light
during inflation,
generating a stochastically independent set of
potential wells for each light
degree of freedom in the gravitational sector.
The implications of this scenario
for the assembly of cosmic structure
have not yet been worked out in detail.

\para{Synopsis}%
In this paper we study
models of modified gravity with an extra
scalar degree of freedom, and
show that if inflation seeded
a set of potential wells for this field
they can renormalize the effective
force law driving mass assembly.
In addition, under certain circumstances
they generate stochastic effects which
could be visible in a sufficiently detailed galaxy survey.
We show that constraints
can be obtained from measurements of the
`two-dimensional' power spectrum $P(k,\mu)$,
which incorporates information about both
local densities and velocities
as a function of redshift.

The paper is divided into three principal parts.
In~\S\ref{sec:structureform}
we generalize the analysis of
density--density,
density--velocity and velocity--velocity
clustering statistics to models
with extra forces and potential wells.
We show that these statistics can be used
to learn about the effective cosmological-scale
force law and the number of dynamically relevant
combinations of potentials.

In~\S\ref{sec:galileon}
we describe a concrete model of this type,
the cubic `Galileon' scenario of Nicolis et al.~\cite{Nicolis:2008in}.
We specialize the discussion of
cosmological mass assembly from~\S\ref{sec:structureform}
and address some of the issues which arise when
deciding whether this model is a viable alternative to
a simple cosmological constant.

In~\S\ref{sec:likelihood}
we explain how the cubic Galileon model can be
compared to a particular dataset
constraining the velocity and density clustering
statistics:
the multipole power spectra measured by the
WiggleZ Dark Energy Survey.
We report the results in~\S\ref{sec:results},
and explain our conclusions in~\S\ref{sec:conclusions}.

We have tried to include sufficient explanation to make
this paper relevant to both theorists and observers.
Readers already familiar with measurements of the growth factor
may wish to omit~\S\ref{sec:structureform} as far
as~\S\ref{sec:modified-gravities}.
The earlier part
is a review of standard material.
In~\S\ref{sec:galileon} we have tried to
summarize the motivation and relevance of `Galileon'
models of modified gravity, and to explain their
current theoretical status.
However, readers who are primarily interested in
constraints on the model itself
may prefer to skip most of~\S\ref{sec:galileon}
on a first reading.

\para{Notation}%
Except when reporting observations,
we work in natural units where $c = \hbar = 1$.
We express the strength of the gravitational force
in terms of the reduced Planck mass,
$\Mp = (8\pi \GNewton)^{-1/2}$, where
$\GNewton$ is the conventional Newton constant.
In our units, $\Mp$ has dimensions of
energy with numerical value
$\Mp = 2.435 \times 10^{18} \, \GeV$.
We measure the strength
of fifth forces in energy units using a
coupling scale $\Coupling$ which is analogous to $\Mp$.

\section{Structure formation and stochastic bias}
\label{sec:structureform}

In this section we review the
analysis of structure formation in
Einstein gravity (\S\ref{sec:einstein-gravity}),
assuming all perturbations to be seeded by a single
primordial fluctuation,
and show how it can be regarded as
a test of the gravitational force law
$\vect{F}/m = - G M \hat{\vect{r}} / r^2$.
In {\S\ref{sec:modified-gravities}} we
generalize to the case
of several independent primordial fluctuations.

\para{Accretion onto overdensities}%
In a statistically homogeneous and
isotropic relativistic cosmological model
we must describe gravity using two
independent potentials $\Phi(\vect{x},t)$ and $\Psi(\vect{x},t)$.
To first order in fluctuations, the metric is
\begin{equation}
    \d s^2 = - (1 + 2 \Psi) \d t^2 + a^2 (1 + 2 \Phi) \d \vect{x}^2 .
\end{equation}
The scale factor $a$ is a function only of time $t$.
In Einstein gravity,
at this order,
the potentials
are equal up to a sign
in the absence of anisotropic stress.
We are assuming that the prehistory of the observable universe
has selected the flat Friedmann--Robertson--Walker
metric, perhaps
as a result of an inflationary era.

At early times the universe consists of
matter and radiation.
The background densities are $\rho_m$, $\rho_r$
and we assume
small perturbations $\delta \rho_m \equiv \delta_m \rho_m$,
$\delta\rho_r \equiv \delta_r \rho_r$.
The spatial configuration of each fluid
is rearranged by
flow fields or `velocity perturbations'
$\vect{v}_m$,
$\vect{v}_r$.
In the potential flow approximation
we can represent these perturbations
by the
divergence $\theta = \vect{\grad}\cdot\vect{v}$.%
	\footnote{Our velocities are in comoving coordinates,
	and $\vect{\grad}_i=\partial/\partial x^i$.}

In a Newtonian model we would obtain
a dynamical equation by
balancing the sum of forces $\sum_i \vect{F}_i$
on a small element of fluid.
Expressing this balance equation in
terms of $\theta$,
allowing for an expanding universe and
the possibility of a modified gravitational force law,%
    \footnote{This appears as a modification
    to the Poisson equation which determines
    $\Psi$. See Eq.~\eqref{eq:poisson-constraint}.}
we find
\begin{subequations}
\begin{align}
    \label{eq:full-thetam-eq}
    \frac{\d \theta_m}{\d t}
    & =
        - 2 H \theta_m
        - \frac{\partial^2}{a^2} \Psi
        + F_{5m} \\
    \label{eq:full-thetar-eq}
    \frac{\d \theta_r}{\d t}
    & =
        - H \theta_r
        - \frac{\partial^2}{a^2} \Psi
        - \frac{1}{4} \frac{\partial^2}{a^2} \delta_r
        + F_{5r}.
\end{align}
\end{subequations}
Here, $-\partial \Psi$ is the static
gravitational force,
terms proportional to $H\theta$ represent Hubble friction, and
$F_{5m}$, $F_{5r}$ describe
(the divergence of) possible fifth-forces acting on
matter and radiation,
respectively. They are zero in pure Einstein gravity.
The term proportional to
$\delta_r$ in Eq.~\eqref{eq:full-thetar-eq}
represents an additional force
due to the internal pressure of radiation.

Eqs.~\eqref{eq:full-thetam-eq}--\eqref{eq:full-thetar-eq}
are dynamical. They show that linear growth of
cosmological structure is essentially an
infall experiment
accounting for work done against the expansion $H$.
To close this system requires further
kinematic equations describing
the deposition of excess radiation or matter into each
fluid element. These are
\begin{subequations}
\begin{align}
    \label{eq:full-deltam-eq}
    \frac{\d \delta_m}{\d t}
    & =
        - \theta_m
        - 3 \dot{\Phi}
        + j_{5m} \\
    \label{eq:full-deltar-eq}
    \frac{\d \delta_r}{\d t}
    & =
        - \frac{4}{3} \theta_r
        - 4 \dot{\Phi}
        + j_{5r} ,
\end{align}
\end{subequations}
where $j_{5m}$, $j_{5r}$ represent matter and radiation fluxes
induced by fifth-forces.
The mass-energy deposited by the flows $\vect{v}_m$, $\vect{v}_r$
is described
by the terms proportional to $\theta_m$, $\theta_r$,
whereas the terms proportional to $\dot{\Phi}$
describe
changes in density due to
relativistic
modulation
of the proper volume by $\Phi$. This approach,
where deviations from the predictions of Einstein gravity
are modelled as fifth forces and fluxes, was
employed in Ref.~\cite{Burrage:2016myt}.

\subsection{Einstein gravity}
\label{sec:einstein-gravity}
In Einstein gravity there are no fifth forces or fluxes,
and
on subhorizon scales the modulation
due to $\dot{\Phi}$ is negligible
and can be discarded.
This is
the `quasi-static' approximation.
In this limit
changes in density are entirely
due to deposition by the flow,
and the deposition equation for matter can be rewritten
as a definition of the \emph{growth factor} $f$,
\begin{equation}
    \label{eq:growth}
    \frac{\d \ln \delta_m}{\d \ln a} \equiv f
    = - \frac{\theta_m}{H\delta_m} .
\end{equation}

When reporting the results of
velocity surveys
it is conventional to
measure $\theta$
in Hubble units, corresponding
to the rescaling $\theta \mapsto \theta / H$.
In what follows we
normalize $\theta$ in this way,
after which~\eqref{eq:growth} becomes
$\theta_m = - f \delta_m$.

In this model the growth rate is spatially independent,
and therefore~\eqref{eq:growth}
implies
that $\theta_m$ is
totally correlated
with $\delta_m$ up to a bias $-f$
depending on the rate of deposition.
According to
Eq.~\eqref{eq:full-thetam-eq}
this deposition is maintained purely
by gravitational
forces.
As we describe
in Eq.~\eqref{eq:f-force-law} and below,
this makes
the precise value of $f$
depend on details
of the gravitational force law.

In Einstein gravity the force law is
determined by the Poisson constraint
\begin{equation}
    \label{eq:poisson-constraint}
    {- \frac{k^2}{a^2}} \Psi = \frac{\rho}{2 \Mp^2}
    \left(
        \delta_m
        + 3H \frac{a^2}{k^2} \theta_m
    \right)
\end{equation}
together with the `no-slip' condition
$\Phi = - \Psi$.
The strength of the force is measured by
the Planck mass $\Mp = (8\pi \GNewton)^{-1/2}$.
On subhorizon scales
the Poisson constraint
expresses
the familiar inverse-square law
$\vect{F}/m = - \GNewton M \hat{\vect{r}}/r^2$.
In modified gravity,
changes to the $1/r^2$
force law are sometimes
expressed as a scale- or redshift-dependent
$\GNewton$ or Planck mass $\Mp$,
\emph{viz.}
$\Mp^2(k) \equiv \beta(k) (\Mp^2)_0$
where $\beta(k) \rightarrow 1$
as $k \rightarrow 0$
and
$(\Mp^2)_0$ represents the strength of
gravity on very large scales.
Then $\beta(k)$ represents a correction
to the $1/r^2$ law on smaller scales.%
    \footnote{A similar parametrization
    introduces an effective rescaling
    $1+\mu(k,z)$ on the right-hand side of the Poisson
    equation.
    Some constraints on $\mu$ exist from data;
    see, for example, Ref.~\cite{Simpson:2012ra,Dossett:2015nda}.
    Any parametrization of this kind
    which absorbs fifth forces
    into a renormalization of the effective
    Planck mass
    assumes
    total correlation between
    all forces.
    It will not apply in the models to be
    described below where
    the correlation is incomplete.}

This parametrization of
a scale-dependent force law
is conventional in the
modified-gravity community.
Note that $\beta(k)$
defined in this way
is unconnected with the ratio of $f$
to the bias, $f/b$,
also denoted $\beta$ within the
large-scale structure community.

\para{Example: change to $1/r^2$ law}%
Combining Eqs.~\eqref{eq:full-thetam-eq},
\eqref{eq:full-deltam-eq}
and~\eqref{eq:poisson-constraint}
yields the M\'{e}sz\'{a}ros equation.
Allowing for scale-dependent modifications
to the force law but
assuming the Friedmann equation is sensitive only to
the long-wavelength force,
the M\'{e}sz\'{a}ros equation can be written
(assuming an $\Omega_m=1$ cosmology)
\begin{equation}
  \ddot{\delta}_m + \frac{4}{3t} \dot{\delta}_m -
  \frac{2\beta(k)^{-1}}{3t^2} \delta_m = 0 .
\end{equation}
The corresponding scale-dependent growth rate is
\begin{equation}
  f(k) = \frac{-1 + \sqrt{1 + 24\beta(k)^{-1}}}{4} .
  \label{eq:f-force-law}
\end{equation}
If $\beta$ decreases then the effective gravitational force
increases, as does $f$.
If $\beta$ increases, the reverse is true.

Given measurements of $f(k)$ it is possible
to recover $\beta(k)$, from which the behaviour of
the force law could be extracted via a Fourier transform.
The outcome
is that
combined probes of the velocity and density
perturbations serve a similar
purpose to laboratory tests of the inverse-square
law, but on vastly different scales.

In this $\Omega_m=1$ model,
if $f=1$ the force law is precisely proportional to
$1/r^2$. Deviations from $f=1$ signal deviations from
inverse-square-law behaviour. Current data favour
$f \sim 0.5$, which can be interpreted
as a softening
on cosmological scales due to the repulsion of
dark energy counteracting the attraction of
Newtonian gravity. A more detailed discussion
of the effect of dark energy perturbations was given
by Nesseris \& Sapone~\cite{Nesseris:2015fqa}.

\subsection{Modified gravities}
\label{sec:modified-gravities}
In more complex models the interpretation of $f$
becomes less clear, because we must separate the
effects of scale- \emph{and} time-dependence in
the force law
from work done against the expansion.
We do not know in advance what should be
attributed to any of these sources.
Nevertheless, provided only a single set of primordial
potential wells exist,
$f$ can be regarded as a measure of the effective
gravitational force including cosmological effects.
Under these circumstances
it is a deterministic function, independent of
position, making $\delta_m$ and $\theta_m$
completely correlated as above.

In models with fifth-forces
it is possible for
inflation to seed more than
one stochastically independent set
of potential wells.
As we explain below,
the local growth rate
can be influenced by contributions
from each set
and
we must regard $f$ as a probe
not only of the force law but also of
the initial conditions.
In this section
our aim
is to set up a framework which is sufficiently
general to describe models
of this kind.

\para{Single set of potential wells}%
If $f$ is not a fixed number but varies from place to
place, we speak of
\emph{stochastic bias}.
Where stochastic effects are important,
the predictions of a physical model become statistical
statements about the expected value or distribution
of observables.
Therefore the discussion should be framed
in terms of correlation functions
rather than quantities such as
$\delta$, $\theta$ or $\Phi$ which are not directly measurable.

First, we replicate the discussion of Einstein gravity
in this language,
assuming there is
only a single
primordial fluctuation which is responsible for sourcing
the large-scale matter distribution and its flow,
normally ascribed to
the Newtonian potential $\Phi$.
We suppose it has
attained
a practically time-independent value $\Phi^\ast(\vect{x})$
by some time early in the radiation era.%
    \footnote{It is a matter of convention to which
    field we ascribe the primordial perturbation.
    In this model the initial conditions for
    $\delta$, $\theta$ and $\Phi$
    are related by constraints.

    \noindent
    If $\Phi^\ast$ has an inflationary origin,
    it may itself be a composite
    of the vacuum fluctuations in the active light
    fields of the model.
    Its precise composition does not become
    fixed until the dynamics have become
    adiabatic.
    We assume this happens sometime before the onset
    of radiation domination.}
Then, at some later time $t$,
the matter and flow fields
can be written
\begin{subequations}
\begin{align}
    \delta_m(k, t) & = \transfer{\Phi}{\delta}(k, t) \Phi^\ast(k) \\
    \theta_m(k, t) & = \transfer{\Phi}{\theta}(k, t) \Phi^\ast(k) ,
\end{align}
\end{subequations}
where we have assumed statistical isotropy
and
$\transfer{j}{i}$ is a transfer function describing
how an initial fluctuation in field $j$ is reprocessed
into a later configuration of field $i$.
In this subsection $\delta$ and $\theta$
always refer to the matter density and velocity,
and we drop the distinguishing subscript `$m$'.

In analogy with the definition
$-f = \theta / \delta$
we define an effective (but possibly
scale- and time-dependent)
bias by
\begin{equation}
    \label{eq:feff-definition}
    \feff(k) \equiv
    - \frac{\transfer{\Phi}{\theta}(k)}{\transfer{\Phi}{\delta}(k)}
    = - \frac{\theta(k)}{\delta(k)} ,
\end{equation}
in which the stochastic initial condition
has been `divided out'.
In this equation and what follows
we suppress the dependence of
$\feff$ and
$\transfer{j}{i}$ on time.

Still assuming only a single primordial
perturbation,
$\feff(k)$
could equivalently be defined
using the relative normalization
of the different two-point
correlation functions,
\begin{subequations}
\begin{align}
    \label{eq:deltathetacross}
    \langle \delta \theta \rangle_k
    & =
        - \feff(k) \langle \delta \delta \rangle_k \\
    \label{eq:thetatheta}
    \langle \theta \theta \rangle_k
    & =
        \feff(k)^2 \langle \delta \delta \rangle_k ,
\end{align}
\end{subequations}
Eqs.~\eqref{eq:deltathetacross}--\eqref{eq:thetatheta}
deal with stochasticity by averaging over it,
rather than dividing it out as in~\eqref{eq:feff-definition}.

Eqs.~\eqref{eq:deltathetacross}--\eqref{eq:thetatheta}
characterize
`deterministic' bias,
which is associated with a fixed multiplicative
normalization
between
$\langle \delta \delta \rangle$,
$\langle \delta \theta \rangle$
and
$\langle \theta \theta \rangle$.
This fixed change in normalization
is the signature of complete correlation between
$\theta$ and $\delta$.
Attempts have been made to detect
this bias experimentally~\cite{Blake:2011rj}.

\para{Multiple sets of potential wells}%
We now add a fifth force,
mediated by a
scalar field $\phi$
which we also assume to have received
a primordial inflationary
fluctuation $\delta\phi^\ast(k)$.
We suppose
that $\delta\phi^\ast(k)$ achieves a
time-independent value by the onset of radiation domination
in the same way as the gravitational potential.
If $\phi$ is active during inflation
this could
leave $\Phi^\ast$ and $\delta\phi^\ast$ partially
or entirely correlated.
However, simple models
can produce the uncorrelated
case $\langle \Phi \delta\phi \rangle^\ast = 0$
and in this paper we focus on scenarios of that type.

Even if
the primordial fluctuations
$\Phi^\ast$ and $\delta\phi^\ast$
are uncorrelated they
will mix in
the late-time matter and flow fields.
Physically this corresponds to a competition between
gravity and the new fifth force to attract
matter into their respective potential wells.
Therefore we must now write
\begin{subequations}
\begin{align}
    \label{eq:delta-m-transfer-general}
    \delta(k) & = \transfer{\Phi}{\delta}(k) \Phi^\ast(k)
    	+ \transfer{\phi}{\delta}(k) \delta\phi^\ast(k) \\
    \label{eq:theta-m-transfer-general}
	\theta(k) & = \transfer{\Phi}{\theta}(k) \Phi^\ast(k)
		+ \transfer{\phi}{\theta}(k) \delta\phi^\ast(k) .
\end{align}
\end{subequations}
We define growth factors
associated with $\Phi$ and $\delta \phi$,
\begin{subequations}
\begin{align}
	\feffNewton & = -\frac{{T_\theta}^\Phi}{{T_\delta}^\Phi}
	\label{eq:feffNewton-def}
	\\
	\feffGalileon & = -\frac	{{T_\theta}^\phi}{{T_\delta}^{\phi}} .
	\label{eq:feffGalileon-def}
\end{align}
\end{subequations}
These
characterize the strength and scale dependence
of the force law associated with infall into each type
of potential well.
The relationship between the $\langle \delta \delta \rangle$,
$\langle \theta \delta \rangle$ and $\langle \theta \theta \rangle$
correlation functions becomes
\begin{subequations}
\begin{align}	
    \label{eq:deltatheta-cross-decorrelated}
    \langle \theta \delta \rangle_k
    & =
    	\langle \delta \delta \rangle_k
    	\bigg(
    		{-\feffNewton(k)} + \frac{\feffNewton(k) - \feffGalileon(k)}{1 + \rho(k)}
    	\bigg)
    \\
    \label{eq:thetatheta-decorrelated}
    \langle \theta \theta \rangle_k
    & =
    	\langle \theta \delta \rangle_k
    	\bigg(
    		{-\feffNewton(k)} + \frac{\feffNewton(k) - \feffGalileon(k)}{1 + \sigma(k)}
    	\bigg)
    \\
    \label{eq:thetatheta-decorrelated-dd}
    & =
    	\langle \delta \delta \rangle_k
    	\bigg(
    		\feffNewton(k)^2 + \frac{\feffGalileon(k)^2 - \feffNewton(k)^2}{1 + \rho(k)}
    	\bigg) .
\end{align}
\end{subequations}
These equations
assume the primordial decorrelation condition
$\langle \Phi \delta \phi \rangle^\ast = 0$
and would require modification were it to be abandoned.

The clustering power
is given
by summing
the contribution
from each set of potential wells.
The same is true for the velocity power.
In Eqs.~\eqref{eq:deltatheta-cross-decorrelated}--\eqref{eq:thetatheta-decorrelated-dd}
we have introduced parameters
$\rho$ and $\sigma$
which
measure the relative contributions to
the $\delta\delta$ and $\theta\delta$
correlation functions
from each set of potentials.
They satisfy
\begin{subequations}
\begin{align}
	\rho(k) & \equiv
		\frac{({T_{\delta}}^\Phi)^2}{({T_{\delta}}^\phi)^2}
		\frac{\langle \Phi \Phi \rangle^\ast}{\langle \delta \phi \delta \phi \rangle^\ast}
	\label{eq:rho-def}
	\\
	\sigma(k) & \equiv
		\frac{{T_{\theta}}^\Phi {T_{\delta}}^\Phi}{{T_{\theta}}^\phi {T_{\delta}}^\phi}
		\frac{\langle \Phi \Phi \rangle^\ast}{\langle\delta\phi\delta\phi\rangle^\ast}
	\label{eq:sigma-def}
\end{align}
\end{subequations}
If the clustering power
is dominated by accumulation within the $\Phi^\ast$
potential wells then $\rho \gg 1$,
whereas if it is dominated by
accumulation within the $\delta\phi^\ast$
potential wells then $\rho \ll 1$.
A similar statement holds for $\sigma$.

Eqs.~\eqref{eq:deltatheta-cross-decorrelated}--\eqref{eq:thetatheta-decorrelated-dd}
show that,
depending which set of potential wells dominate
the correlation functions,
their relative normalization
interpolates between
$\feffNewton$ and $\feffGalileon$.
Where only Newtonian potential wells are relevant
the effective growth factor is $\feffNewton$,
and where only Galileon potential wells are relevant
the effective growth factor is $\feffGalileon$.

In a typical model
there are two possibilities.
First, matter may accumulate
within just one of the primordial potentials,
perhaps $\Phi^\ast$.
Taking this as an example,
the fifth force
will still be relevant if it modifies the background expansion
history or its perturbations couple to $\Phi$.
Either possibility can introduce scale- or time-dependent
modifications of the Newtonian force law, leading to changes
in $\feffNewton$---%
but
the growth factors measured
from different combinations of the two-point functions will agree.
The net effect is a Galileon analogue of the `softening'
from $f=1$ to $f \sim 0.5$ due to dark energy
in a $\LCDM$ model.
This scenario was studied by
Appleby~\& Linder in the case where
the fifth-force field $\phi$ has Galileon
interactions.
They found it to be incompatible with observation
if the Galileon made a significant contribution to the
expansion history~\cite{Appleby:2011aa,Appleby:2012ba}.

Alternatively,
matter may accumulate in a combination
of the $\Phi^\ast$ and $\delta\phi^\ast$
potential wells.
The net force driving inflow into this combination
will be an admixture of the force
laws associated with each type of potential well.
As we explain below, in the most favourable circumstances
we may be able to
observe different combinations of these laws
using different combinations of the two-point functions
such as
$-\langle \theta \delta \rangle / \langle \delta \delta \rangle$
and
$-\langle \theta \theta \rangle / \langle \theta \delta \rangle$.
In this scenario we could measure at least two
different effective growth factors.
Further combinations become available
where more than one fifth-force field and corresponding
set of potential wells
exist.
A qualitatively similiar discussion applies if
$\Phi^\ast$ and $\delta\phi^\ast$ are correlated,
although modified in detail because the force laws
no longer operate independently.

This `decorrelation'
occurs only if
$\sigma(k)$ is appreciably different to $\rho(k)$,
so that the power in
$\langle \delta \delta \rangle$
and $\langle \theta \delta \rangle$
is dominated by clustering around different
combinations
of the potential wells.
For example, this might happen if most matter
has accumulated in one combination of potentials
but a strong flow is driving exchange with the
orthogonal
combination.
Such decorrelations
are the signature of
nondeterministic or \emph{stochastic} bias.
If it occurs the signature is unambiguous
because
the relationship between
the two-point functions cannot be mimicked by
\emph{any} choice of deterministic bias,
even one which is scale-dependent.
The deterministic model implies
that different ways to measure the
growth factor must agree,
because they are always measuring the same force law.
Specifically,
\begin{equation}
	\frac{\langle \theta \theta \rangle}
	{\langle \theta \delta \rangle}
	-
	\frac{\langle \theta \delta \rangle}
	{\langle \delta \delta \rangle}
	=
	\Big(
		\feffNewton - \feffGalileon
	\Big)
	\frac{\rho - \sigma}{(1+\rho)(1+\sigma)} ,
\end{equation}
and the right-hand side is zero
in a deterministic model
irrespective of the force law
and the primordial power spectrum.
Whether departures from zero are
measurable depends on $\rho$, $\sigma$ and the
split between $\feffNewton$ and $\feffGalileon$.
In \S\S\ref{sec:likelihood}--\ref{sec:results}
we will study this sort of
decorrelation in a concrete Galileon
model.

A systematic study of the relative
magnitude of
$\langle \delta \delta \rangle$,
$\langle \delta \theta \rangle$
and
$\langle \theta \theta \rangle$
as functions of scale therefore
yields two outcomes.
First,
we probe the effective strength
and scale-dependence of the gravitational force law.
Second,
we determine whether
a single combination of potential wells
was relevant during structure formation,
or whether there is evidence for a more complex model.

Similar effects can occur for any pair of
fluctuations.
For example, in certain models of modified gravity,
the first-order relation $\Phi = - \Psi$ is violated
(described as `slip').
This can be probed through combinations
of observables that measure both
metric potentials, such as weak gravitational
lensing or the integrated Sachs--Wolfe effect.
Stochastic effects would generate systematic
shifts between the correlation functions
of $\delta$ with (for example)
the weak-lensing shear $\gamma$,
or the ISW
temperature decrement,
that are not simply fixed changes in normalization.
The effectiveness of such combinations for
constraints on non-minimally coupled modified-gravity models
was studied by Gleyzes et al., although the analysis is limited
to the quasi-static approximation~\cite{Gleyzes:2015rua}.

Valkenburg \& Hu have made available a code for 
generating initial conditions for cosmological N-body simulations, 
which accommodates perturbation spectra that are decorrelated or 
anticorrelated between the matter and extra scalar degree of freedom~\cite{ValkenburgHu}.

\section{Galileon modified gravity}
\label{sec:galileon}

In the remainder of this paper we apply the formalism
developed in \S\ref{sec:structureform}
to a specific model for the fifth-force
field $\phi$,
which we take to be a Galileon~\cite{Nicolis:2008in}.
We do not make use of the quasi-static approximation.
In~\S\ref{sec:galileon-model}
we briefly recall the relevance of the
model for scenarios of modified gravity,
and in~\S\ref{sec:background-expansion}
we discuss the background cosmological solution and
its expansion history.
In~\S\ref{sec:galileon-structure}
we prepare for the calculation of effective
growth factors by specializing the structure-formation
equations from~\S\ref{sec:structureform}.
Finally, in~\S\ref{sec:galileon-vacuum-flucts}
we discuss the criteria which should be used to
decide whether the model suffers from
objections similar to those which led us to
reject a simple cosmological constant.

\subsection{The Galileon model}
\label{sec:galileon-model}

A Galileon is a
normal scalar field whose self-interactions are
restricted to be of a special kind~\cite{Deffayet:2009wt,Deffayet:2009mn}.
The action for a Galileon coupled to gravity
is of the form
\begin{equation}
    S = \int \d^4 x \; \sqrt{-g}
    \Big(
        \frac{\Mp^2}{2} R
        - \frac{1}{2} \sum_{2 \leq i \leq 5} c_i \SelfInteraction^{3(2-i)} \Lag_i ,
    \Big)
\end{equation}
where $g_{ab}$ is the metric,
$R$ is the Ricci scalar built from
$g_{ab}$ and its Levi--Civita connexion,
and the allowed $\Lag_i$ are
\begin{subequations}
\begin{align}
    \label{eq:L-galileon-2}
    \Lag_2 & = ( \nabla \phi )^2 \\
    \label{eq:L-galileon-3}
    \Lag_3 & = ( \nabla \phi )^2 \Box \phi \\
    \label{eq:L-galileon-4}
    \Lag_4 & = ( \nabla \phi )^2 \Big[
            2 ( \Box \phi )^2
            - 2 \phi_{;\mu\nu} \phi^{;\mu\nu}
            - \frac{R}{2} ( \nabla \phi )^2
        \Big] \\
    \nonumber
    \Lag_5 & = ( \nabla \phi )^2 \Big[
        ( \Box \phi )^3
        - 3 ( \Box \phi ) \phi_{;\mu\nu} \phi^{;\mu\nu} \\
    \label{eq:L-galileon-5}
        & \qquad \qquad \hspace{2mm}
        \mbox{} + 2 {\phi_{;\mu}}^\nu {\phi_{;\nu}}^\rho {\phi_{;\rho}}^\mu
        - 6 G_{\nu \rho} \phi_{;\mu} \phi^{;\mu\nu} \phi^{;\rho}
    \Big] .
\end{align}
\end{subequations}
A semicolon denotes the covariant derivative
compatible with $g_{ab}$ and
$G_{\nu\rho}$ is the Einstein tensor.
We have also used $(\grad \phi)^2 = g^{ab} \phi_{;a} \phi_{;b}$
and $\Box \phi = g^{ab} \phi_{;ab}$.
If there are no hierarchies in the Galileon sector
it will be possible to choose $\SelfInteraction$
so that the $c_i$ are of order unity.
Then $\SelfInteraction$ determines where
the higher-order interactions $\Lag_3$, $\Lag_4$
and $\Lag_5$ become comparable
to $\Lag_2$.

The Galileon model is an interesting example
in which to
test probes of modified gravity.
We adopt it for several reasons:

\sidehead{1. Screening through the Vainshtein mechanism}%
    Although $\phi$ mediates a fifth force,
    it may be dynamically suppressed
    (`screened') sufficiently close to
    heavy objects
    due to
    nonlinear effects associated with the
    scale $\SelfInteraction$~\cite{Nicolis:2008in,Burrage:2010rs}.
    Dealing correctly with such nonlinear
    screening effects
    is expected to be important for
    realistic scenarios.

    Several distinct screening mechanisms
    exist~\cite{Joyce:2014kja}. Galileons exhibit a type known as
    `Vainshtein' screening, in which
    suppression of the fifth force occurs rather
    softly and over large distances.
    The WiggleZ survey, which we will use for
    comparison to data, is particuarly
    suitable for tests of Vainshtein-type screening
    because most objects in the survey are
    field galaxies far from cluster centres.
    This offers an opportunity to probe
    the transition from screened to unscreened
    fifth forces around cluster outskirts.

    In addition to screening of sources,
    there can be `cosmological' screening
    which universally
    suppresses fifth forces when the mean
    cosmological energy density is large.
    The Galileon model exhibits this
    `cosmological' Vainshtein effect~\cite{Chow:2009fm}.

\sidehead{2. Versatile interpretation}%
    Galileon models contain no physics which
    could solve the
    cosmological constant problem,
    but it is possible
    that they approximate the long-wavelength
    behaviour of other (currently unknown)
    physics which does. In
    particular:
    \begin{enumerate}
        \item If the
        effective cosmological constant
        is somehow set to zero,
        late-time acceleration could be
        supported by an energy density associated
        with $\phi$.
        As explained in~\S\ref{sec:structureform},
        this possibility has been considered by
        Appleby \& Linder
        and later authors~\cite{Appleby:2011aa,Appleby:2012ba,Barreira:2012kk,Okada:2012mn,Barreira:2013jma,Barreira:2013eea},
        who found that it led to
        changes in the growth of cosmic structure
        which were excluded by observation.
        Other constraints on Galileon models which do not explicitly
        couple to matter have been obtained by Barreira et al.
        using observations of the cosmic microwave background and
        baryon acoustic oscillations~\cite{Barreira:2014jha}.

        \item Alternatively, $\phi$ may have no role
        in supporting late-time acceleration
        but merely be a vestige of whatever physics
        is responsible for it.
        For example, this might happen in
        a massive gravity where the graviton
        mass is responsible for `degravitating' the
        cosmological constant.
        Then $\phi$ is an unavoidable by-product of the
        graviton mass but is not otherwise important.
        Galileon models have been shown to arise
        in this way from the de Rham--Gabadadze--Tolley
        massive gravity~\cite{deRham:2010ik,deRham:2010kj}.
        In this scenario $\phi$ has negligible energy
        density.
    \end{enumerate}
For the second possibility there are two relevant questions:
first, whether expansion histories exist in which the
Galileon is always subdominant;
and second, whether its contribution to the effective
gravitational force nevertheless excludes the model.
In this paper we focus on scenarios of this type.

\sidehead{3. Sufficiently concrete}%
    The model is sufficiently well-defined
    that we can (and should)
    apply an analysis
    similar to the logic which
    led us to reject vacuum fluctuations as an
    explanation for the microphysics of acceleration.
    This is important because nothing is gained
    by replacing a simple model with a more
    complex one which suffers from the same drawbacks.
    We discuss this issue in \S\ref{sec:galileon-vacuum-flucts}.

\subsection{Cosmological background expansion}
\label{sec:background-expansion}

We focus on the cubic Galileon
model, for which only $c_2$ and $c_3$ are nonzero.
This is numerically simplest and already includes
the interesting features of the $\Lag_4$ and $\Lag_5$
terms, except for anisotropic stress which
(to first-order in perturbations)
is
identically zero
for $\Lag_3$.

\para{Matter species}%
We take the matter to consist of
radiation and pressureless dust representing
cold dark matter,
which we couple
to $\phi$ via the conformal transformation
\begin{equation}
    \label{eq:conformal-coupling}
    g_{\mu\nu} \mapsto
    \Big(
        1 + \frac{\phi}{\Coupling}
    \Big)
    g_{\mu\nu}
\end{equation}
within the matter Lagrangian.
This means that $\phi$ couples to
the trace of the energy--momentum tensor
$T = {T^a}_a$
for the matter fields via an interaction of the
form $\phi T / \Coupling$.
For a mix of matter and radiation
the trace $T$ equals the matter density
$\rho_m$.

When measured using the Einstein-frame
metric $g_{\mu\nu}$
this conformal coupling
makes the matter density and pressure
$\phi$-dependent.
They are related to their $\phi$-independent
Jordan-frame counterparts
$\rho^J$, $p^J$
(which can be regarded
as the `intrinsic' density and pressure for the
fluid) by the rule
\begin{subequations}
\begin{align}
    \rho & =
        \Big(
            1 + \frac{\phi}{\Coupling}
        \Big)
        \rho^J \\
    p & =
        \Big(
            1 + \frac{\phi}{\Coupling}
        \Big)
        p^J .
\end{align}
\end{subequations}
The $\phi$-dependence allows the Einstein-frame
fluid to exchange energy with the Galileon,
described by the
continuity equations
\begin{subequations}
\begin{align}
        \label{eq:matter-continuity}
    \frac{\d \rho_m}{\d t} + 3 H \rho_m
        & =
        - \frac{1}{2\Coupling}
        \frac{\d \phi}{\d t}
        \Big(
            1 + \frac{\phi}{\Coupling}
        \Big)^{-1}
        \rho_m \\
    \label{eq:radiation-continuity}
    \frac{\d \rho_r}{\d t} + 4 H \rho_r & = 0 ,
\end{align}
\end{subequations}
where $t$ is the Einstein-frame time, $H$ is
the corresponding Hubble parameter, and
$\rho_m$ and $\rho_r$ are, respectively, the
energy density in CDM and radiation.
Eq.~\eqref{eq:radiation-continuity} shows that
the radiation is conserved separately, without exchanging
energy with any other component.
This happens because the action
for radiation is classically conformally invariant.

\para{Galileon evolution}%
The Galileon field profile is controlled
by the field equation
\begin{equation}
    \label{eq:galileon-field-equation}
	c_2 \Box \phi
	+ \frac{c_3}{\SelfInteraction^3}
	    [
	    (\Box\phi)^2-\partial_{\mu}\partial_{\nu}\phi\partial^{\mu}\partial^{\nu}\phi
	    ]
	- \frac{1}{2 \Coupling} \Big( 1 + \frac{\phi}{\Coupling} \Big)^{-1}
	T = 0,
\end{equation}
If they are required,
expressions for the contribution of each higher-order
Galileon operator to the equation 
of motion can be found in Refs.~\cite{Nicolis:2008in,Deffayet:2009wt}.
We include all nonlinear effects in $\phi$
because these
are important to correctly capture the
`cosmological' Vainshtein effect,
described in more detail below.

Specialising to the background cosmological evolution, Eq.~\eqref{eq:galileon-field-equation} can be expressed
as a continuity equation for $\phi$,
\begin{equation}
	\frac{\d \rho_\phi}{\d t} + 3 H (\rho_\phi + p_\phi) =
		\frac{\dot{\phi}}{2\Coupling}
		\Big( 1 + \frac{\phi}{\Coupling} \Big)^{-1} \rho_m .
	\label{eq:phi-conservation}
\end{equation}
It can be checked that~\eqref{eq:matter-continuity}, \eqref{eq:radiation-continuity}
and~\eqref{eq:phi-conservation}
together imply total conservation of energy.

\para{Cosmological Vainshtein effect}%
We are focusing on scenarios where the
Galileon remains a subdominant
contributor to the universe's expansion history.
To achieve this we must select a suitable
trajectory for the
background $\phi$ field,
which we describe as the
`cosmological Vainshtein solution'. 
We seek a solution in which
nonlinear effects dominate the
field profile.
The background Galileon configuration
evolves according to
\begin{equation}
    c_2 ( \ddot{\phi} + 3 H \dot{\phi} )
    + \frac{c_3}{\SelfInteraction^3} \Big[
        \frac{\d}{\d t} ( H \dot{\phi}^2 )
        + 3 H^2 \dot{\phi}^2
    \Big]
    =
        - \frac{\rho_m}{2\Coupling} .
\end{equation}
Assuming the $c_3$ term is dominant
and that $H\sim 1/t$,
which is the case during matter and radiation domination, then
\begin{equation}
    \dot{\phi} \sim \frac{\SelfInteraction}{H}
    \sqrt{-\frac{\rho_m \SelfInteraction}{c_3 \Coupling}} .
    \label{eq:cosmological-vainshtein-solution}
\end{equation}
We must choose $\SelfInteraction > 0$ to give
$\phi$ a stable kinetic term, so
this `cosmological Vainshtein solution'
exists only if $\Coupling < 0$.
The same conclusion was reached by
Chow \& Khoury, who studied
cosmological evolution in a number
of Galileon models~\cite{Chow:2009fm}.
Note that the sign of $\Coupling$ has no effect on the
nature of the force mediated between matter particles
by $\phi$. This depends on $\Coupling^2$ and is always
attractive.

On the solution~\eqref{eq:cosmological-vainshtein-solution}
the Galileon energy density relative to matter
can be written
\begin{equation}
    \frac{\rho_\phi}{\rho_m}
    =
    - \frac{1}{c_3^{1/3}}
    \frac{\Mp}{\SelfInteraction} \left(
        \frac{3\SelfInteraction}{2|\Coupling|}
    \right)^{3/2}
    \left(
        \frac{\rho_m}{\rho_m + \rho_r}
    \right)^{1/2} .
    \label{eq:relative-galileon-energy}
\end{equation}
Therefore $\rho_\phi$ is negative,
leading to catastrophic consequences if
the Galileon energy density becomes a significant
fraction of the total energy budget.
This is a pathology peculiar to the cubic Galileon model.

We intend~\eqref{eq:relative-galileon-energy} to
apply for a strongly self-interacting
model where
$\SelfInteraction \ll |\Coupling| \lesssim \Mp$.
With these choices it is possible
for the Galileon energy
density
to remain small compared to the matter energy
density
for much of the lifetime of the universe,
and the problem can be avoided.
The precise Galileon contribution is controlled by the
relative abundance of matter and radiation.

The cosmological Vainshtein solution
is an attractor in the space of
solutions~\cite{Appleby:2011aa,DeFelice:2011aa}.
It is valid provided
\begin{equation}
    \frac{c_2^2}{c_3} \SelfInteraction^3 |\Coupling| \lesssim \rho_m .
    \label{eq:cosmological-vainshtein-validity}
\end{equation}
The left-hand side is fixed once we have chosen
values for $c_2$, $c_3$, $\SelfInteraction$ and
$\Coupling$.
These should be selected so that~\eqref{eq:cosmological-vainshtein-validity}
is satisfied at early times.
However, as the
matter density dilutes~\eqref{eq:cosmological-vainshtein-validity}
must eventually be invalidated,
causing the Galileon to contribute an increasing
fraction of the total energy budget.
In the cubic model this leads to a singularity
because of~\eqref{eq:relative-galileon-energy}.
If we
demand that~\eqref{eq:cosmological-vainshtein-validity}
is valid until today, in order to obtain a
nearly $\LCDM$-like expansion history
down to redshift $z \sim 0$,
we obtain
a rough constraint $\SelfInteraction \lesssim 10^{-13}(\Mp/|M|)^{1/3} \, \eV$.

As a side effect, the cosmological Vainshtein solution
suppresses Galileon fifth forces
while the average cosmological energy density
is sufficiently large. This will be discussed further in
\S\ref{sec:screening-linear-validity}.

\subsection{Perturbations and structure formation}
\label{sec:galileon-structure}

Perturbations in the Galileon model can be described using the
general formalism assembled in~\S\ref{sec:structureform}.

\para{Governing equations}%
The dynamical and deposition equations
are~\eqref{eq:full-thetam-eq}--\eqref{eq:full-thetar-eq}
and~\eqref{eq:full-deltam-eq}--\eqref{eq:full-deltar-eq},
with the fifth forces and fluxes
\begin{subequations}
\begin{align}
    \label{eq:matter-fifth-force}
    F_{5m} & = \frac{1}{2\Coupling}
        \Big(
            1 + \frac{\phi}{\Coupling}
        \Big)^{-1}
        \Big(
			\dot{\phi} \theta_m
			+ \frac{\partial^2}{a^2} \delta \phi
		\Big) \\
	\label{eq:matter-fifth-flux}
	j_{5m} & = - \frac{\delta\dot{\phi}}{2\Coupling}
	    \Big(
	        1 + \frac{\phi}{\Coupling}
	    \Big)^{-1}
		+ \frac{\dot{\phi}}{2\Coupling} \frac{\delta \phi}{\Coupling}
		\Big(
		    1 + \frac{\phi}{\Coupling}
		\Big)^{-2} \\
	\label{eq:radiation-fifth-flux}
	F_{5r} & = j_{5r} = 0 .
\end{align}
\end{subequations}
The fifth-force
divergence $F_{5m}$
which sources the matter flow
consists of a potential-gradient contribution
proportional to
$\partial^2 \phi$
and a reaction term
proportional to $\dot{\phi} \theta_m$.
The reaction term arises from
changes in momentum
due to
conversion of matter into $\phi$.
The flux
$j_{5m}$
represents the effective deposition of matter
due to the reverse process,
in which energy density from $\phi$
is locally converted into matter.
The radiation is not coupled to $\phi$
and is therefore unaffected, giving
$F_{5r} = j_{5r} = 0$.

We also
require an evolution equation for the
Galileon perturbation $\delta\phi$,
which can be written
\begin{equation}
	D_2 \delta\ddot{\phi} + \cdots =
		\frac{\rho_m \delta_m}{2\Coupling}
		\Big( 1 + \frac{\phi}{\Coupling} \Big)^{-1}
		- \frac{\rho_m}{2\Coupling} \frac{\delta \phi}{\Coupling}
		\Big( 1 + \frac{\phi}{\Coupling} \Big)^{-2} ,
    \label{eq:galpert5}
\end{equation}
where $D_2 = -c_2 +6c_3 H \dot{\phi}/\SelfInteraction^3$
is a coefficient depending on the background cosmology
and Galileon evolution,
and
`$\cdots$' denotes a series of terms that are first-order
in perturbations, all of which enter with coefficients
that are functions of the background quantities.
Full expressions for all these coefficients are given by
de Felice, Kase \& Tsujikawa~\cite{DeFelice:2010as}.

\para{Initial conditions}%
We set initial conditions in the radiation era.
The Galileon field velocity is chosen to select
the cosmological Vainshtein
solution~\eqref{eq:cosmological-vainshtein-solution}.

The density and velocity perturbations inherit their
initial amplitudes from the
primordial Newtonian and Galileon potentials.
In the cubic Galileon model there is no anisotropic stress
to first order, and therefore at this order
the Einstein equations
require $\Phi = -\Psi$.
We work in terms of $\Phi$.
On superhorizon scales,
deep in the radiation era where both
matter and Galileon densities are
negligible, the initial condition is
\begin{equation}
    \delta_r^\ast(\vect{k}) = 2 \Phi^\ast(\vect{k}) ,
\end{equation}
where a superscript `$\ast$' denotes evaluation at the initial time.

As explained in {\S\ref{sec:structureform}},
we are taking $\Phi^\ast$ and $\delta\phi^\ast$
to be stochastic random fields
whose statistical properties are determined
by a prior inflationary phase,
and we are assuming that
$\langle \Phi \delta\phi \rangle^\ast = 0$.
We take the $\Phi^\ast$
two-point function to satisfy
\begin{equation}
    \langle \Phi(\vect{k}_1) \Phi(\vect{k}_2) \rangle^\ast
    = (2\pi)^3 \delta(\vect{k}_1 + \vect{k}_2)
    \frac{\AmplitudeNewton}{k^3}
    \Big(
        \frac{k}{\kpiv}
    \Big)^{n_s-1} ,
    \label{eq:Phi-twopf}
\end{equation}
where $\AmplitudeNewton$ is an amplitude
to be fixed using Planck measurements of the microwave
background power spectrum, and $n_s$ is
a spectral index.
We also assume that during inflation
$\delta\phi$ was light
enough to
receive
a quantum fluctuation
and write its two-point function
\begin{equation}
    \langle \delta \phi(\vect{k}_1) \delta \phi(\vect{k}_2) \rangle^\ast
    = (2\pi)^3 \delta(\vect{k}_1 + \vect{k}_2)
    \frac{\AmplitudeGalileon}{k^3} .
    \label{eq:deltaphi-power-spectrum}
\end{equation}
We define the amplitude
$\AmplitudeGalileon$,
and likewise $\AmplitudeNewton$ and $n_s$,
at the Planck pivot scale $\kpiv = 0.05 \, \Mpc^{-1}$.
In this paper we assume the $\delta\phi$ fluctuations
do not have a significant scale dependence.
We use the Planck2013+WP best-fit value
$n_s = 0.9619$ and assume no running~\cite{Ade:2013zuv}.

To fix the amplitudes $\AmplitudeNewton$ and $\AmplitudeGalileon$
we use the Planck2013+WP best-fit value
for the amplitude of
the primordial curvature perturbation
power spectrum,
$A_s = 2.215 \times 10^{-9}$,
to infer the $\Phi$
power spectrum on a large scale $k = 0.001 h/\Mpc$
measured by Planck near $z \sim 1000$
in their best-fit $\LCDM$ cosmology.
We assume a fixed
ratio $\xi$
between the depths of the primordial
Newtonian and Galileon potential
wells on this scale
\begin{equation}
	\langle \delta\phi \delta\phi \rangle^\ast|_{k=0.001 h/\Mpc}
	=
		\xi \Mp^2 \langle \Phi \Phi \rangle^\ast|_{k=0.001 h/\Mpc},
	\label{eq:xi-def}
\end{equation}
and match
to the inferred
large-scale $\Phi$ power spectrum near $z \sim 1000$.
If $\xi \sim 1$ then on large scales the Newtonian and
Galileon potential wells are of comparable depth.

Initial conditions for the remaining perturbations
can be obtained from
equations~\eqref{eq:full-thetam-eq}--\eqref{eq:full-thetar-eq},
\eqref{eq:full-deltam-eq}--\eqref{eq:full-deltar-eq}
and~\eqref{eq:matter-fifth-force}--\eqref{eq:radiation-fifth-flux}.
The matter density contrast satisfies
\begin{equation}
    \delta_m^\ast(\vect{k}) =
        \frac{3}{4} \delta_r^\ast(\vect{k})
        - \frac{\delta\phi^\ast(\vect{k})}{2\Coupling} .
\end{equation}
Causality prevents coherent motion on
superhorizon scales,
so velocity perturbations must be
at least of order $(k/a)^2$.
Imposing that the solutions
are nearly static yields suitable initial conditions,
\begin{subequations}
\begin{align}
    H^\ast \theta_r^\ast(\vect{k}) & \approx
        - \Big(
            \frac{k}{a^\ast}
        \Big)^2
        \frac{\Phi^\ast(\vect{k})}{2} \\
    H^\ast \theta_m^\ast(\vect{k}) & \approx
        - \Big(
            \frac{k}{a^\ast}
        \Big)^2
        \frac{\Phi^\ast(\vect{k})}{2}
        - \frac{1}{4\Coupling}
        \Big(
            \frac{k}{a^\ast}
        \Big)^2
        \delta \phi^\ast(\vect{k}) .
\end{align}
\end{subequations}

\subsection{Vacuum fluctuations in the Galileon model}
\label{sec:galileon-vacuum-flucts}
Finally, if we intend to replace a simple cosmological constant
by the Galileon model described in this section
(or any other model), we should address whether
it exhibits the same issues which led us to reject
vacuum fluctuations as a viable explanation for
the acceleration.
The Galileon model is sufficiently concrete that we can carry
out part of this check in some detail.

To realize an acceptable expansion
history
requires a choice for the parameters
$\SelfInteraction$ and $c_i$ and the field
trajectory $\phi(t)$.
The Galileon model is no more acceptable
than the model of vacuum energy
if this combination of $\SelfInteraction$
and $c_i$ would be destroyed by quantum fluctuations
around
the required trajectory.

\para{Nonrenormalization}%
Although the Galileon operators in
Eqs.~\eqref{eq:L-galileon-2}--\eqref{eq:L-galileon-5} involve higher time-%
derivatives of $\phi$, these cancel out in the equations of motion.
When quantized this corresponds to the absence of ghosts,
which would otherwise destabilize the vacuum.

In the Minkowski vacuum, quantum fluctuations controlled
by the
operators~\eqref{eq:L-galileon-2}--\eqref{eq:L-galileon-5}
are massless and do not cause renormalization-group
running~\cite{Nicolis:2008in,Nicolis:2004qq,Luty:2003vm}.
Therefore a consistent ghost-free theory exists.
It is currently unclear whether a regime can exist in which
this ghost-free theory is an effective description of
a more complete model that describes interactions
at very high energies~\cite{Kaloper:2014vqa,deRham:2013hsa}.
Even if it does,
there is further uncertainly regarding
survival of the ghost-free property
when heavy sources generate a nontrivial background for $\phi$.
This is required for a consistent Vainshtein mechanism
(including the cosmological Vainshtein effect
described above),
because this relies on renormalization
of the kinetic term for fluctuations in the vicinity
of heavy sources. These renormalizations are generated by
time- or space-gradients of the background field.

In the presence of a nontrivial background, the mass of
$\phi$ fluctuations becomes field-dependent
and calculations become more difficult.
In a cosmological context
the mass terms contain curvature quantities
associated with the background.
Because they are no longer massless, these fluctuations
can induce running of operators
such as $(\Box \phi)^2$
which satisfy the
same symmetries as~\eqref{eq:L-galileon-2}--\eqref{eq:L-galileon-5}
but induce a ghost and therefore destabilize the vacuum.
To our knowledge there is not yet an estimate of this running.
In this paper we assume that it is negligible,
and that a regime exists in which a Vainshtein mechanism
can operate without destabilization.
If the Galileon model eventually turns out not
to satisfy this condition, it would invalidate the
detailed analysis presented below but not
the general principles we are describing.

\para{Matter coupling}%
These issues are particular problems of principle
associated with the Galileon model,
although very similar considerations will arise
in any field theory where higher-dimension
operators become relevant on the background.
In the case of Galileon models more theoretical
work is needed to clarify whether fine-tuning
problems exist comparable to those which afflict
the cosmological constant.

A more tractable class
of problems arise
if $\phi$ is coupled to the Standard Model,
as assumed in~\S\ref{sec:background-expansion},
because of renormalizations due to matter
loops.
These problems (or similar ones)
will also occur in almost any field theory
model which couples extra gravitational
degrees of freedom to matter.

Once we couple $\phi$ to Standard Model matter,
the coupling will typically be renormalized.
For example, restricting attention
to the `conformal'
coupling of Eq.~\eqref{eq:conformal-coupling},
we could have selected a different
rescaling factor
$g_{\mu\nu} \mapsto \Omega(\phi / \Coupling)
g_{\mu\nu}$
for some arbitrary function $\Omega$.
At tree-level we can choose $\Omega$ freely,
but once we have done so
it will be renormalized by
matter loops.
This leads to a loss of predictivity if
we rely on any details of the
form of $\Omega$
as a function of $\phi/\Coupling$.
To maintain radiative stability the
best we can do is require $|\phi/\Coupling| \ll 1$,
for which the approximate form
$\Omega \approx 1 + \phi/\Coupling$
will be preserved under renormalizations.
The scale
$\Coupling$
is the analogue of $\Mp$
for the force mediated by $\phi$ and
must be constrained by observation.

Any coupling to matter breaks the Galileon
symmetry, although only mildly if $|\phi/\Coupling|$
remains small.
Having done so,
matter loops will typically generate
operators such as $(\partial \phi)^4 / \Coupling^4$
which do not not satisfy the
Galilean symmetry
leading to~\eqref{eq:L-galileon-2}--\eqref{eq:L-galileon-5}.
And in any case, we could
have introduced other symmetry-breaking
terms by hand at the same time as
the symmetry-breaking coupling~\eqref{eq:conformal-coupling}.%
    \footnote{In this paper we remain in Einstein frame
    throughout, using the metric $g_{\mu\nu}$.
    However, the issue of symmetry-breaking contributions to
    the Lagrangian
    is complicated
    by the possibility of frame transformations.
    A conformal transformation which makes the matter
    sector minimally coupled
    introduces symmetry breaking operators
    which mix the scales $\SelfInteraction$ and $\Coupling$,
    such as $(\partial \phi)^4 / \SelfInteraction^3 \Coupling$.
    These terms show that a Galileon theory
    specified in Jordan frame by the nonminimal
    gravitational term
    $f(\phi/\Coupling) R/2$
    requires extra symmetry-breaking operators
    if it is to be equivalent to the
    Einstein-frame theory specified by
    the same nonminimal conformal coupling
    to matter.
    While this raises no problem of principle,
    it \emph{does} mean that we must be explicit
    regarding the frame in which the theory is
    defined.
    Without the inclusion of extra symmetry-breaking
    terms, the Jordan- and Einstein-frame theories
    are equivalent only if
    the missing
    terms such as
    $(\partial \phi)^4/\SelfInteraction^3\Coupling$
    are negligible.
    We would like to thank Joseph Elliston for
    detailed discussions regarding this issue.}
We collectively denote all these terms by $\Lbreak$.
The test for~\eqref{eq:conformal-coupling}---or
any other choice---is whether such corrections are
irrelevant compared to those already present
in the Galileon action.

\para{Numerical evolution}
In the numerical work to be described
in~\S\ref{sec:likelihood}
we estimate the importance of
quantum effects by tracking the size of $|\phi / \Coupling|$
and a sample of possible contributions to
$\Lbreak$.

The first of these representative contributions is
the ratio
$(\partial \phi)^2/\Coupling^4$,
which represents the generic magnitude of corrections
to each Galileon operator from (for example)
loops of spin-$\frac{1}{2}$ matter.
It would normally be accompanied by a radiatively
generated mass term whose size we do not attempt
to estimate.

The second contribution is the
ratio $(\partial \phi)^2 / \Coupling \Box \phi$.
This represents the relative importance
of corrections such as $(\partial \phi)^4 / \SelfInteraction^3 \Coupling$
to the cubic Galileon term in the equation of motion.
This term is generated by changing conformal frame,
or could be generated by loops mixing Galileon and
matter fluctuations.

We reject models where any of these
ratios become larger than $10^{-3}$.
This cutoff is intended to keep
corrections
at the sub-percent level.
There is no necessary implication that
models in which it is violated
are a poor fit for the data, but only that
we cannot obtain reliable predictions within
our present framework due to the possibility
of large radiative corrections.
We should therefore exclude them,
because their status is no better than the
cosmological constant model they are
intended to replace.%
    \footnote{It is usually
    assumed that the cosmological constant relevant for
    estimates of
    the Hubble rate should be computed by matching
    (at some energy
    scale $M \gg H_0$
    where $H_0 \approx 10^{-33} \, \eV$
    is the present-day expansion rate~\cite{Weinberg:1988cp})
    a low-energy theory in which the long-wavelength
    gravitational force is described by
    Einstein gravity
    to some other, more complete theory.
    This matching
    prescription is generally accepted
    although it is possible to imagine
    that the correct theory of quantum gravity
    requires a different choice,
    perhaps because the
    cosmological constant has no clear
    physical meaning in the high-energy theory---%
    unlike scattering amplitudes,
    to which such matching calculations
    have been traditionally been applied,
    the cosmological
    constant can apparently only be measured
    (indirectly)
    at very long wavelengths.

    Accepting the conventional prescription,
    we can reliably calculate the running of the
    cosmological constant with the matching scale
    at energies $\ll M$ where the relevant degrees of
    freedom are known from experiment.
    This would give a contribution at least of order
    the top-quark mass to the fourth
    power, $m_t^4 \sim (170 \, \GeV)^4$---%
    some $(10^{14})^4$
    orders of magnitude larger than the
    observed value, which is roughly $(10^{-3} \, \eV)^4$.
    See, eg., Ref.~\cite{Martin:2012bt}.}
Clearly there is some arbitrariness in
deciding what magnitude of corrections
we are prepared to tolerate, and our choice
is somewhat conservative.

As explained above,
we do not attempt to estimate the size of
pure Galileon self-renormalizations
which could introduce a low-scale ghost.
For the $\LCDM$-like background we employ,
we assume these not to be generated
at the matching scale
by the supposed ultraviolet completion.
Whether they are subsequently generated
at a lower scale by radiative corrections involving
Galileon loops
is a complex question which do not attempt to answer here.

\section{Data analysis}
\label{sec:likelihood}

In this section we use data from the WiggleZ
Dark Energy Survey to
constrain the cubic Galileon model of~\S\ref{sec:galileon}.
This gives an explicit example in
which
the effective force driving inflow
of matter
can receive significant scale-dependent
renormalizations due to mixing with
the primordial Galileon potential wells.
We show how a study of the $\langle \delta \delta \rangle$,
$\langle \theta \delta \rangle$
and $\langle \theta \theta \rangle$ correlation functions
constrains such deviations from the effective $\LCDM$ force law.

In~\S\ref{sec:observations} we explain how a galaxy redshift survey
can be used to measure the two-point correlation functions,
in specific combinations called the `multipole power spectra'.
In~\S\ref{sec:screening-linear-validity} we discuss the limits
of validity of the linear analysis presented
in~\S\S\ref{sec:structureform}--\ref{sec:galileon}
and explain the cuts used to restrict our analysis to modes for
which linear theory should be an acceptable approximation.
In~\S\ref{sec:numerics} we give details of our numerical
procedure.
Finally, in~\S\ref{sec:wigglez} we briefly describe the
WiggleZ Dark Energy Survey and explain the computation of
our likelihood function.

\subsection{Measuring $\delta$ and $\theta$ using a galaxy survey}
\label{sec:observations}

Up to this point the discussion has been phrased in
terms of the
$\langle \delta \delta \rangle$,
$\langle \theta \delta \rangle$
and $\langle \theta \theta \rangle$ correlation functions.
In practice we cannot measure these correlation functions
directly.
Rather than $\delta$ we observe the galaxy overdensity
$\delta_g$,
which is related to the density perturbation by an
unknown bias, $\delta_g \approx b \delta$.
Also, velocities are measured using
projection effects and to describe these we must
assume a model.

\para{Redshift-space distortions}%
In this paper we use the
damped Gaussian model
suggested by Peacock~\cite{Peacock1992,Peacock:1993xg}.
The observed $\vect{k}$-space
density contrast is written
\begin{equation}
    \label{eq:kaiser}
    \delta(\vect{k}) \rightarrow
    \deltaobs(\vect{k}) = \e{-\mu^2 \sigma^2 k^2/2H_0^2}
    \big( \delta - \mu^2 \theta \big) ,
\end{equation}
where $\mu = \hat{\vect{x}} \cdot \hat{\vect{k}} \equiv \cos \vartheta$
can be thought of as the cosine of the
angle $\vartheta$ between our line of sight
$\hat{\vect{x}}$
and a wavevector $\vect{k}$
contributing to $\delta$.
The exponential prefactor represents
power suppression
due to virialization on small scales,
which randomizes velocities.
We take virialization to occur for distances smaller than
the scale $\sigma/H_0$
defined by
the pairwise galaxy velocity dispersion $\sigma$
along the line of sight.
Below this scale we can recover
very little information about cosmological-scale
force laws.
When estimating the likelihood
for some particular model we
regard both $\sigma$ and $b$ as parameters
to be fitted.

The Gaussian model provides a reasonable
description,
but on sufficiently small scales
the effects of virialization are complex
and it must be replaced by
a more sophisticated model.
The Kaiser formula~\eqref{eq:kaiser}
continues to apply in the modified-gravity
scenarios we are considering,
and we assume the Gaussian model
does likewise.

\begin{figure*}
    \subcaptionbox{Two-dimensional power spectrum with nearly-negligible
    			   Galileon potential wells,
    			   chosen so that
    			   at early times
    			   $\langle \delta \phi \delta \phi \rangle^\ast
    			   \approx 0$
    			   on large scales.
    			   The fluctuations in this model are
    			   indistinguishable from $\LCDM$.
    			   \label{fig:2dps-smooth}}
                  {\includegraphics[scale=0.29]{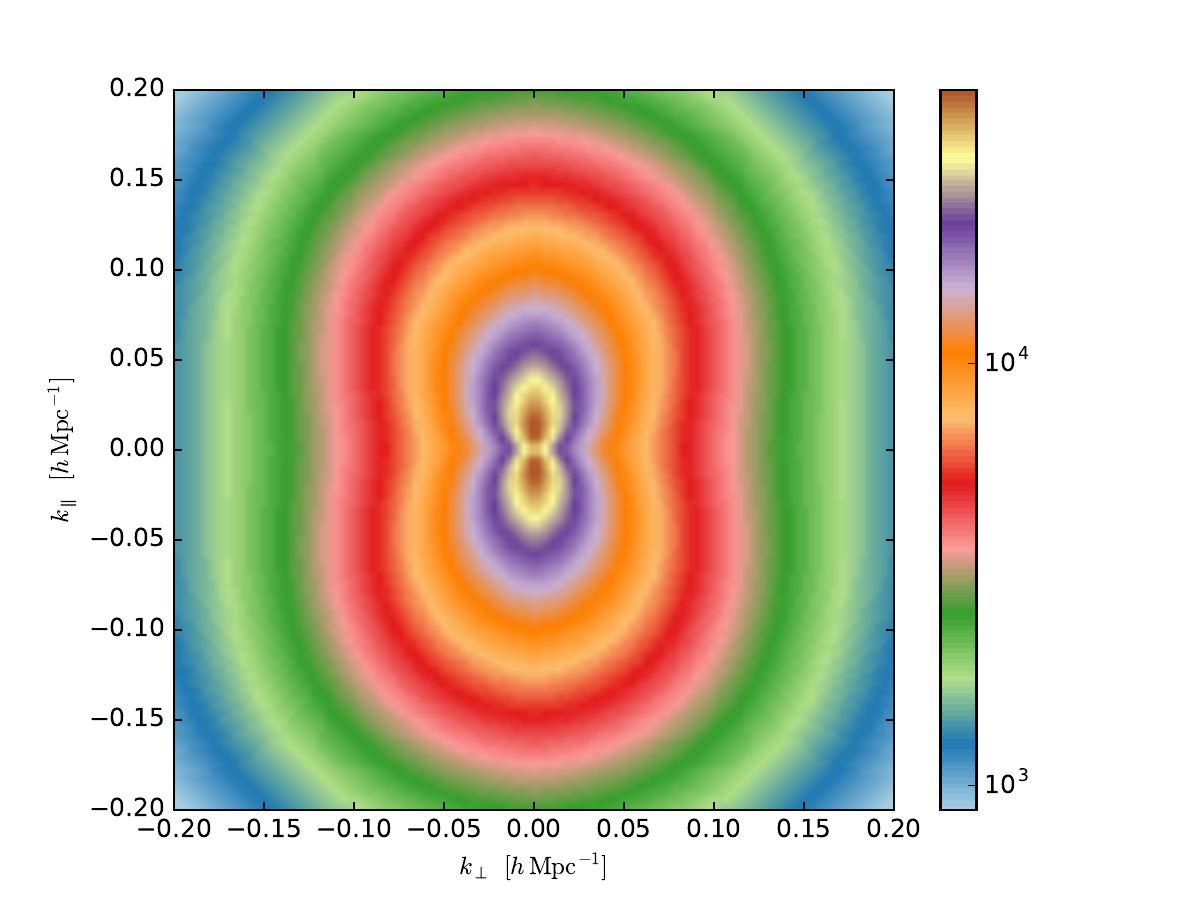}}
    \subcaptionbox{Same background as Fig.~\ref{fig:2dps-smooth},
                   but with more significant Galileon potential wells
                   for which $\langle \delta \phi \delta \phi \rangle^\ast
    			   = 10 \Mp^2 \langle \Phi \Phi \rangle^\ast$ on large scales.\label{fig:2dps-bad}}
                  {\includegraphics[scale=0.29]{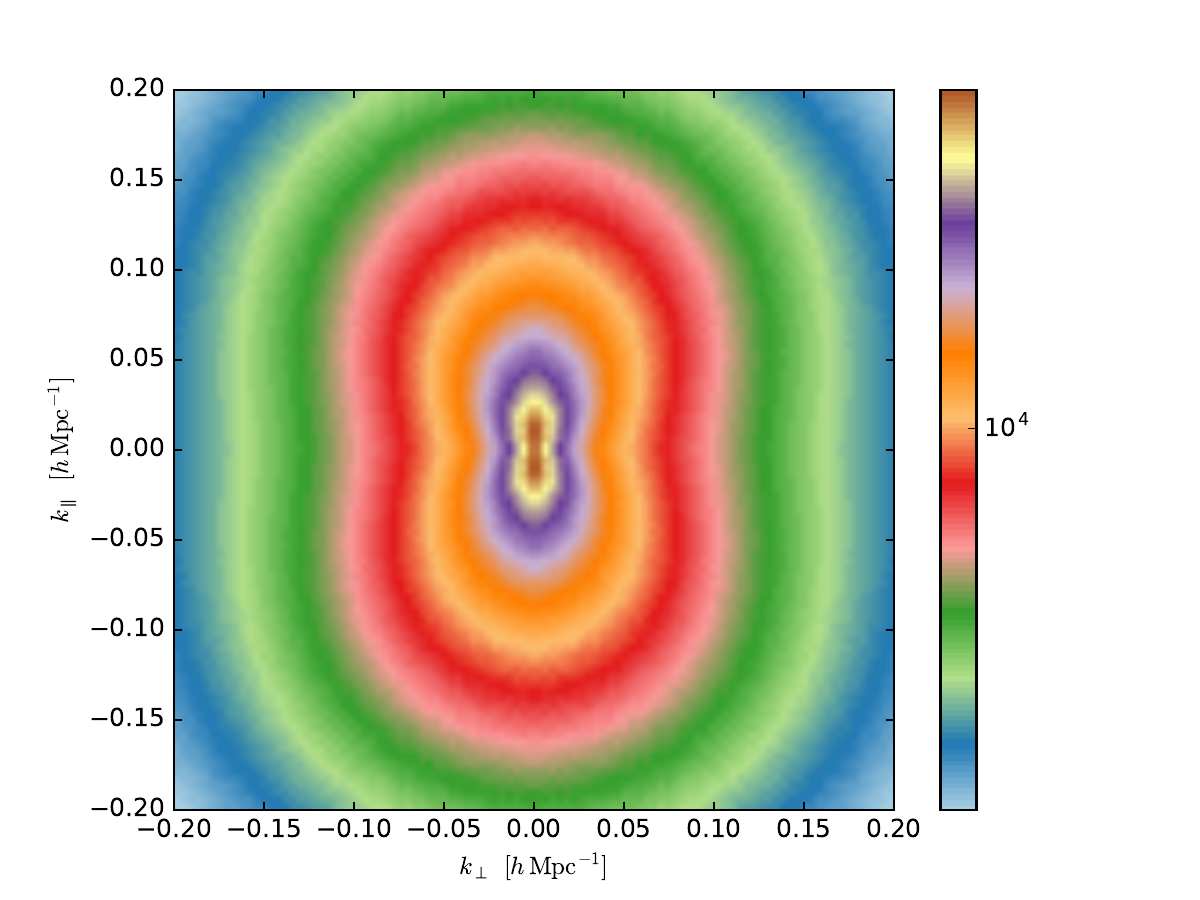}}
    \subcaptionbox{Same background as Fig.~\ref{fig:2dps-smooth},
    			   but with exaggerated
       			   Galileon potential wells
       			   for which
       			   $\langle \delta \phi \delta \phi \rangle^\ast
    			   = 100 \Mp^2 \langle \Phi \Phi \rangle^\ast$ on large scales.\label{fig:2dps-worse}}
                  {\includegraphics[scale=0.29]{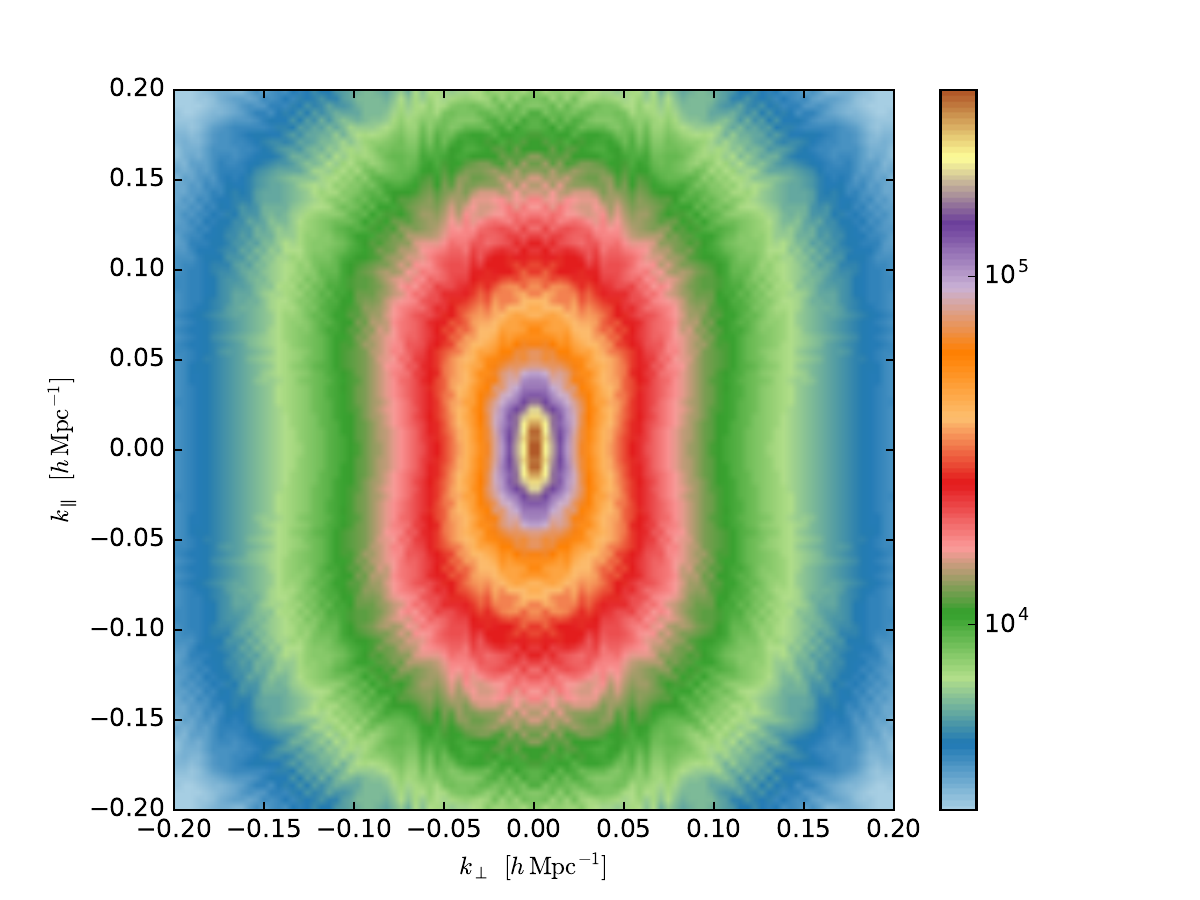}}
    \caption{Examples of two-dimensional power spectra.
    The background cosmology is close to the WiggleZ
    best-fit combination estimated using power-spectrum
    data, with $\Omega_m = 0.28$ and
    $\fbaryon = 0.15$.
    The line-of-sight velocity dispersion is taken to be $\sigma = 320 \, \mathrm{km} \, \mathrm{s}^{-1}$.
    $k_{\perp}$ and $k_{\parallel}$ represent the magnitude of
    momenta perpendicular and parallel to the line of
    sight from earth; ie., $\vect{k}_{\parallel} = k \mu \hat{\vect{x}}$
    and $\vect{k}_{\perp} = \vect{k} - \vect{k}_{\parallel}$.
    Only the top-right quadrant contains independent information; the
    other quadrants are obtained by reflections.\label{fig:2dps}}
\end{figure*}

The observed galaxy-density autocorrelation function is a
$\mu$-dependent combination of the
$\langle \delta \delta \rangle$,
$\langle \delta \theta \rangle$
and $\langle \theta \theta \rangle$
correlation functions,
\begin{equation}
    \langle \delta_g \delta_g \rangle_{\obs}
    =
    \e{-k^2 \mu^2 \sigma^2 / H_0^2}
    \Big(
        b^2 \langle \delta \delta \rangle
        - 2 b \mu^2 \langle \theta \delta \rangle
        + \mu^4 \langle \theta \theta \rangle
    \Big) .
\end{equation}
By fitting for the $\mu$-dependence of the measured
power spectrum we can extract each of the components.
From the correlation function
we define a power spectrum
$P_{\obs}$
by writing
\begin{equation}
    \langle \delta_g(\vect{k}_1) \delta_g(\vect{k}_2) \rangle_{\obs}
    =
        (2\pi)^3 \delta(\vect{k}_1 + \vect{k}_2)
        P_{\obs}(k,\mu) ,
    \label{eq:P-delta-delta-obs}
\end{equation}
In Fig.~\ref{fig:2dps} we plot $P_{\obs}(k,\mu)$
as a function of $\vect{k}_{\parallel}$ and
$\vect{k}_{\perp}$, defined by
$\vect{k}_{\parallel} = k \mu \hat{\vect{x}}$
and $\vect{k}_{\perp} = \vect{k} - \vect{k}_{\parallel}$,
in a Galileon model with varying relative depths of
the primordial potential wells.

In current experiments the angular resolution
is too coarse for this reconstruction to be performed
with high accuracy.
Instead,
Cole, Fisher \& Weinberg
decomposed the angular
dependence of $P_{\obs}$ into multipoles~\cite{Cole:1993kh}.
Taking
$\Legendre{\ell}(\mu)$ to be
the Legendre polynomial of order $\ell$,
the multipoles $P_{\ell}$ satisfy
\begin{equation}
    P_{\obs}(k,\mu)
    \equiv \sum_{\ell} P_{\ell}(k) \Legendre{\ell}(\mu) .
\end{equation}
They can be computed
using orthogonality of the $\Legendre{\ell}$,
\begin{equation}
    P_{\ell}(k) \equiv \frac{2 \ell + 1}{2}
        \int_{-1}^{+1} \d \mu \; P_{\obs}(k,\mu) \Legendre{\ell}(\mu) .
    \label{eq:multipole-def}
\end{equation}
In~{\S\ref{sec:numerics} we will see that this prescription requires
modifications to account for anisotropic scaling due to the Alcock--Paczynski
effect.

Within the Gaussian model
only even multipoles can be nonzero,
because $P_{\obs}$ is a function of $\mu^2$.
The monopole $P_0$ is the angle-averaged power spectrum,
which was used in
the analysis of Ref.~\cite{Parkinson:2012}.

\para{Information content}%
Each multipole blends information from
the
$\langle \delta \delta \rangle$,
$\langle \theta \delta \rangle$
and $\langle \theta \theta \rangle$
power spectra
and therefore contains an admixture
of the information on the force laws
and number of dynamically relevant potentials
carried by these correlation functions.
The relative normalizations
described in~\eqref{eq:deltatheta-cross-decorrelated}--\eqref{eq:thetatheta-decorrelated-dd}
fix the amplitudes of the $P_{\ell}$
as a function of scale.
Also,
because different combinations of
correlation functions contribute to each
multipole at fixed $k$,
decorrelation of
the $\delta$ and $\theta$ two-point functions
will modify the relative normalization of the $P_\ell$.

\subsection{Screening and validity of linear theory}
\label{sec:screening-linear-validity}

The formalism developed in \S\S\ref{sec:structureform}--\ref{sec:galileon}
applies only to linear order in
the fluctuations.
When comparing its predictions to data we must be
careful to exclude any modes for which
nonlinear effects may have been important.

In a $\LCDM$ model
this is not difficult, because at any
redshift there is a fixed scale beyond which
structure has become nonlinear.
In a modified gravity the situation may be
different.
In the Galileon model
there is a `Vainshtein effect',
in which non-linear behaviour  of the
field $\phi$ suppresses fifth-forces
in the vicinity of large mass concentrations.
On a flat background the transition from unscreened to screened
fifth-forces occurs
roughly at the Vainshtein radius~\cite{Vainshtein:1972sx,Nicolis:2008in,Burrage:2010rs}
\begin{equation}
    \RVainshtein \equiv
        \frac{1}{\SelfInteraction}
        \Big(
            \frac{c_3 \Msource}{2\pi c_2^2 |\Coupling|}
        \Big)^{1/3} .
    \label{eq:vainshtein-radius}
\end{equation}
In our calculation we retain $\delta\phi$ only to first order.
Therefore our formalism can not
account for this Vainshtein effect. 
This should not be confused with the cosmological Vainshtein effect discussed in
\S\ref{sec:background-expansion}, which is a property of the background evolution.

Because this scale~\eqref{eq:vainshtein-radius}
depends on the source mass $\Msource$,
the question of whether a fixed physical scale
experiences screened or unscreened fifth forces
throughout the survey volume
depends on the realization of the density fluctuation
within it---in particular, on the
most massive condensation.
Determining the smallest scale on which linear
theory continues to apply then becomes
a probabilistic exercise requiring the methods
of extreme-value statistics.
Even worse, the smallest unscreened scale may
depend on the Lagrangian parameters
and can therefore vary over parameter space.

In the cubic Galileon model these difficulties can be evaded.
Inspection of~\eqref{eq:galpert5}
shows that the cosmological Vainshtein
effect corresponds to the approximate
rescalings
\begin{equation}
    \SelfInteraction \mapsto \SelfInteraction D_2^{1/2}, \quad
    \Coupling \mapsto \Coupling D_2^{1/2}.
\end{equation}
Therefore, on the
background~\eqref{eq:cosmological-vainshtein-solution},
the Vainshtein radius is rescaled by a factor
\begin{equation}
    \RVainshtein \mapsto \RVainshtein / D_2^{2/3}.
\end{equation}
It can be checked that this gives
\begin{equation}
    \RVainshtein^3 =
        \frac{1}{72\pi} \frac{c_3}{c_2^2} \frac{\Msource}{\rho_m}
        \sim \frac{\Msource}{\rho_m} ,
    \label{eq:rescaled-vainshtein}
\end{equation}
which is independent of the self-interaction
scale $\SelfInteraction$
and the coupling $\Coupling$.
Assuming $c_2 \sim c_3$ so that there are no
hierarchies in the Galileon sector,
it depends only on the source mass and
the background matter density.
This conclusion would be spoiled in a more general
Galileon model
where the cosmological and conventional
Vainshtein effects scale
differently with
$\SelfInteraction$ and $\Coupling$
if they are controlled by different
$\Lag_i$.

For a source of density $\rhosource$ and radius $\Rsource$,
Eq.~\eqref{eq:rescaled-vainshtein}
gives
\begin{equation}
    \RVainshtein = \left( \frac{1}{54} \frac{c_3}{c_2^2} \frac{\rhosource}{\rho_m} \right)^{1/3}
        \Rsource
        \sim \left( \frac{\rhosource}{\rho_m} \right)^{1/3} \Rsource ,
\end{equation}
and therefore $\RVainshtein \sim \Rsource$ unless $\rhosource$
is very different from the cosmological average $\rho_m$.
On cluster scales
the density concentration is not very large
and we can expect the linearized analysis
to apply on all scales which are safely larger
than the clusters themselves. We sketch
this situation in Fig.~\ref{fig:vainshtein}.

\begin{figure}
    \includegraphics[scale=0.8]{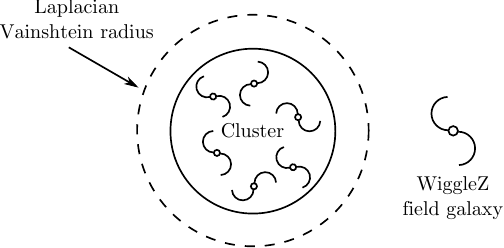}
    \caption{\label{fig:vainshtein}Galaxy pairs
    separated by short distances
    (eg. within a dense cluster environment) are screened by the
    Vainshtein mechanism. Galaxies with
    larger separations (eg. distant cluster members
    interacting with the central condensation) are unscreened, and therefore
    experience the effects of the fifth forces.
    In the WiggleZ survey most galaxies are of the second kind.}
\end{figure}

In practice we use data for wavenumbers
$k \leq 0.2 h/\Mpc$
corresponding to a smallest physical scale of
roughly $31h^{-1} \, \Mpc$. An alternative proposal,
where the calculated power spectrum is inverse-weighted by 
local density to measure an `unscreened power spectrum', has
recently been suggested by Lombriser, Simpson \& Mead~\cite{Lombriser:2015axa}.

\subsection{Numerical procedure}
\label{sec:numerics}

\para{Baryonic effects}%
We evolve Eqs.~\eqref{eq:full-thetam-eq}--\eqref{eq:full-thetar-eq},
\eqref{eq:full-deltam-eq}--\eqref{eq:full-deltar-eq}
and~\eqref{eq:matter-fifth-force}--\eqref{eq:radiation-fifth-flux}
for different wavenumbers
to construct the transfer functions $\transfer{i}{j}$
defined in~\S\ref{sec:modified-gravities}.
These equations do not account for a separate baryonic
component,
which would require a numerical Boltzmann code
including the effect of Galileon fluctuations.
Instead, we model the effect
of baryon suppression
in the density contrast using a version of the
fitting formulae
given by
Eisenstein \& Hu~\cite{Eisenstein:1997ik, Eisenstein:1997jh}.
This is implemented by constructing new transfer functions
$\mathsf{T}(k) = T(k/\Gammaeff)$,
where
$\Gammaeff$ is defined by Eq.~(16) of
Ref.~\cite{Eisenstein:1997jh}.
We neglect the density of neutrinos
and take
the epoch of matter--radiation equality and
the drag epoch to be those measured by Planck,
\emph{viz.}
$\zeq = 3403$ and $\zdrag = 1059.25$.
The baryon fraction is fixed to be $\fbaryon = 0.15$.
We do not include the effect of baryon acoustic oscillations.

Eisenstein \& Hu's formulae were calibrated by comparison
to a Boltzmann code which assumed
a $\LCDM$ cosmology.
They are unlikely to
remain quantitatively accurate
if the Galileon density constitutes a
nonnegligible fraction of the energy budget
of the universe
or if Galileon forces become very significant.

\para{Growth rates}%
To accurately predict the normalization of the
multipoles $P_{\ell}$
we require a reliable
estimate of the relative relationship between
$\delta$ and $\theta$.
We assume that the linear-theory calculation gives a reliable
prediction for the growth rates
associated with infall into the Newtonian and Galileon
potential wells,
given by $\feffNewton$ and $\feffGalileon$
[see Eqs.~\eqref{eq:feffNewton-def}--\eqref{eq:feffGalileon-def}].
We estimate the velocity transfer functions
including baryonic effects
using
${\mathsf{T}_\theta}^\Phi = \feffNewton {\mathsf{T}_\delta}^\Phi$
and
${\mathsf{T}_\theta}^{\delta\phi} = \feffGalileon {\mathsf{T}_\delta}^{\delta\phi}$,
where
${\mathsf{T}_\delta}^\Phi$ and ${\mathsf{T}_\delta}^{\delta\phi}$
are the Eisenstein \& Hu transfer functions discussed above.
The modified transfer functions
${\mathsf{T}_\delta}^\Phi$,
${\mathsf{T}_\delta}^{\delta\phi}$,
${\mathsf{T}_\theta}^\Phi$ and
${\mathsf{T}_\theta}^{\delta\phi}$
are used to produce our final power spectra.

\para{CMB spectrum}%
We do not use microwave background measurements except to fix the normalizations
$\AmplitudeNewton$ and $\AmplitudeGalileon$.
In particular we do not attempt to compute the microwave
background angular power spectrum.
This could receive corrections if the Galileon
fluctuations are significant around the time of last scattering,
but we leave a detailed analysis for future work.

\subsection{WiggleZ Dark Energy Survey}
\label{sec:wigglez}

The WiggleZ Dark Energy Survey at the
Australian Astronomical Observatory
was designed to extend the study of large-scale structure
to
redshifts $z > 0.5$, complementing SDSS observations at lower redshifts.
The survey began in August 2006 and completed observations in January 2011.
It has obtained of order 200,000 redshifts for
UV-bright emission-line galaxies, covering close to
1000 square degrees of equatorial sky.
The design of the survey and its selection function
are described in Ref.~\cite{Drinkwater:2010a}.

\para{Multipole power spectra}%
The first three multipole power spectra
(the monopole $P_0$, quadrupole $P_2$ and hexadecapole $P_4$)
have been determined in three overlapping redshift
bins, $0.2 < z < 0.6$, $0.4 < z< 0.8$ and $0.6 < z < 1.0$,
and for six separated areas of the sky.
These correspond to right ascensions of 1-hr, 3-hr, 9-hr, 11-hr, 15-hr and 22-hr
on the celestial sphere, matching the regions observed by the
survey.
The effective redshifts of these bins are $\zeff=[0.43, 0.6, 0.78]$.

In this section we show that these multipole power spectra
can be used to obtain constraints on the matter coupling $\Coupling$
and depth of the primordial potential wells in the cubic Galileon model.
Our numerical procedure, described in~{\S}\ref{sec:numerics},
is not exact and therefore our constraints will not be optimal.
We intend them as a proof of principle.
A better analysis could be performed with suitable inclusion
of baryonic effects and modelling of the microwave background power spectrum.

We use only the lowest and highest redshift bins for which the
redshift ranges are separate and non-overlapping.
Correlation of bins between different redshifts and
regions is assumed to be negligible.
However, correlations between the multipole moments
in the same bin are important.
The covariance matrix for the three multipole power spectra
has been estimated
using mock catalogues generated by the Comoving Lagrangian
approach~\cite{Tassev:2013pn}
in a similar way to those described in Ref.~\cite{WiggleZBAO}.
Full details of these
simulations are given in Ref.~\cite{Koda:ip}.

\para{Alcock--Paczynski effect}%
To convert redshift-space measurements of galaxy positions
into an estimated power spectrum requires a fiducial
cosmological model
with which to translate observed angles and redshifts
into distances. When comparing the power spectrum predicted by a given model
to the observed power spectrum we must account for
discrepancies between the model and this fiducial cosmology.
This may introduce artificial anisotropies that confuse
the extraction of redshift-space distortions.
To account for this effect we introduce anisotropic scaling
factors
$a_\perp$, $a_\parallel$
that satisfy~\cite{Tegmark:2006,Reid:2009,Swanson:2010}
\begin{subequations}
\begin{align}
    a_\perp & \equiv D_A(z) / \hat{D}_Z(z) \\
    a_\parallel & \equiv \hat{H}(z) / H(z) ,
\end{align}    
\end{subequations}
where $D_A$ is the angular diameter distance
in the model, and $H$ is its Hubble rate;
the hatted quantities $\hat{D}_A$ and $\hat{H}$
are the corresponding values in the fiducial model.
Ballinger et al. showed that
we should rescale $k$ and $\mu$
to new values $k'$ and $\mu'$~\cite{Ballinger:1996cd},
\begin{subequations}
\begin{align}
    k' & \equiv \frac{k}{a_\perp}
        \bigg[
            1 + \mu^2 \left( \frac{a_\perp^2}{a_\parallel^2} - 1 \right)^{1/2}
        \bigg]
        \\
    \mu' & \equiv \frac{\mu a_\perp}{a_\parallel}
        \bigg[
            1 + \mu^2 \left( \frac{a_\perp^2}{a_\parallel^2} - 1 \right)^{-1/2}
        \bigg] .
\end{align}
\end{subequations}
The multipole power spectra defined in Eq.~\eqref{eq:multipole-def}
should be computed using $k'$ and $\mu'$,
\begin{equation}
    P_{\ell}(k) \equiv \frac{2\ell + 1}{2 a_\perp^2 a_\parallel}
    \int_{-1}^{1} \d \mu \; P_{\obs}(k', \mu') \Legendre{\ell}(\mu) .    
\end{equation}

Even after this rescaling has taken place
we must account for incompleteness in the survey volume
due to the observing strategy.
This requires convolution with the survey window function.
For example, incompleteness may be caused
by the presence of bright stars which obscure
galaxies.
If uncorrected these holes
can introduce fake large-scale
structure into the survey.
Including the effect of the window function
yields
\begin{equation}
    \label{eq:alcock-paczynski}
    \Pcon_{\ell_i} (k_i, \zeff) \equiv
    \sum_j W_{ij}(\zeff)
    P_{\ell_j} \big[ k_j, \zeff \big] ,
\end{equation}
where
$\Pcon_{\ell}(k_i, \zeff)$
is the measured multipole power spectrum,
and
the indices $i$ and $j$ run over
all combinations of multipoles $\ell$
and wavenumbers $k$
measured by the survey.
The matrix
$W_{ij}$ encodes details of the window function.

\para{Model likelihood}%
The final likelihood $\Lik$ is
calculated by summing over regions,
redshift bins, and multipole power spectra,
\begin{equation}
    -2\ln \Lik =
        \sum_{\substack{\text{redshifts $z$} \\ \text{regions $r$}}}
        \discrepancy_{r,z}^\transpose
        \Cov^{-1}_{r,z}
        \discrepancy_{r,z} ,
\end{equation}
where $\discrepancy_{r,z}$
is the difference between the
measured and predicted
values of the multipole power
spectra,
\begin{equation}
    \discrepancy_{r,z} \equiv
    \Pmodel_z - \Pdata_{r,z} .
    \label{eq:discrepancy-def}
\end{equation}
In this equation,
$\Pmodel_z$ is a vector of the
multipole power spectra $(P_0, P_2, P_4)$
predicted at redshift $z$
by the model in question---%
rescaled and evaluated at shifted wavenumbers
as described in~\eqref{eq:alcock-paczynski};
$\Pdata_{r,z}$
is a similar vector of multipole power spectra
estimated from the survey in
region $r$ and redshift bin $z$;
and
$\Cov_{r,z}$ is the covariance matrix
for $r$ and $z$,
accounting for
covariances between the different multipole power spectra.
The region label $r$ runs over all regions in the survey,
and $z$ runs over all redshift bins.

\para{Determination of $b$ and $\sigma$}%
The undetermined parameters are the bias $b$,
which may be a function of redshift,
and the line-of-sight velocity dispersion $\sigma$.
We determine $b$ by performing a maximum-likelihood
estimate in each redshift bin,
allowing $b$ to vary between $0.4$ and $1.8$.
We determine $\sigma$ in a similar way,
although it is not taken to be redshift dependent.
We allow it to vary between
$50 \, \mathrm{km} \, \mathrm{s}^{-1}$
and
$650 \, \mathrm{km} \, \mathrm{s}^{-1}$.
The best-fit values of the bias typically depend
on $\sigma$.

\section{Results}
\label{sec:results}

In Fig.~\ref{fig:likelihood}
we plot $1\sigma$, $2\sigma$ and $3\sigma$
best-fit regions in the $(\Omega_m, \Coupling)$
plane
for a range of models with
different
relative depths of the Newtonian
and Galileon potential wells.

\para{Self-interaction scale}%
We choose $\SelfInteraction=1.05 \times 10^{16} H_0$
which enables a `cosmological Vainshtein' solution
to exist [see~\eqref{eq:cosmological-vainshtein-solution}];
on this solution,
our results are nearly independent of
$\SelfInteraction$ at fixed $\Coupling$.
The principal effect of varying $\SelfInteraction$
is to decrease the acceptable range of $\Coupling$.
Because increasing $\SelfInteraction$
implies that~\eqref{eq:cosmological-vainshtein-validity}
can be invalidated more easily,
the Galileon
departs more quickly from the cosmological Vainshtein solution.
After departure, the quantity $|\phi/\Coupling|$ which determines
the importance of quantum corrections to the matter coupling
typically becomes of order unity,
signalling that the model
of~\S\ref{sec:galileon}
becomes untrustworthy due to
uncontrollable loop corrections.
Therefore, as $\SelfInteraction$ increases the range of
$\Coupling$ which maintain $|\phi/\Coupling| \ll 1$ throughout
the evolution becomes smaller.

For values of $|\Coupling| \lesssim \Mp$
(which covers the entire region plotted
in Figs.~\ref{fig:likelihood-neg}, \ref{fig:likelihood-lo},
\ref{fig:likelihood-hi}, and~\ref{fig:likelihood-eq}),
the \emph{linear} Galileon fifth-force is stronger than gravity.
One might have expected this region to be immediately excluded
due to excessive fifth forces.
However, one must remember the cosmological Vainshtein
mechanism: this keeps the background close to
$\LCDM$ and suppresses Galileon forces until the mean cosmic
energy density is sufficiently low.
If $|\Coupling/\Mp|$ is not too small,
this means that fifth-force effects do not become strong
until comparatively late.

\para{Dependence on $\xi$}%
In Fig.~\ref{fig:likelihood-neg}
we set $\xi \approx 0$, making
the primordial Galileon potentials
negligible. With this choice the
initial conditions are indistinguishable
from $\LCDM$.
The best fit region is centred on $\Omega_m \sim 0.25$,
with the coupling constrained to
satisfy
$|\Coupling| \gtrsim 5\times 10^{-3} \Mp$.
This lies just above
the region
(bounded by the blue contour)
where loop corrections become important.
Increasing $\SelfInteraction$
would move the contour almost rigidly
up the plot.
The preference for a slightly low value of $\Omega_m$ is
a feature of the WiggleZ multipole dataset, and is
not connected with our modified-gravity model.

In Fig.~\ref{fig:growth-factors}
we plot the different measures of the growth factor
$-\langle \theta\delta \rangle / \langle \delta\delta \rangle$
and $- \langle \theta\theta \rangle / \langle \theta\delta \rangle$
discussed in \S\ref{sec:modified-gravities},
for each value of $\xi$ appearing in Fig.~\ref{fig:likelihood}.
We measure these
at $z=0.44$
in a model with
$\Omega_m = 0.25$,
$\SelfInteraction = 1.05 \times 10^{16} H_0$
and $\Coupling = -6 \times 10^{-3} \Mp$
(marked by the green diamonds in
Fig.~\ref{fig:likelihood}).
This lies near the bottom of the $1\sigma$
region in Fig.~\ref{fig:likelihood-neg},
and its fit becomes increasingly poor
as $\xi$ increases.

Inspection of Fig.~\ref{fig:growth-factors-neg}
shows that
when $\xi \approx 0$
the velocity bias is precisely deterministic;
there is no evidence for a second set of dynamically
relevant potential wells.
The
constraint on $\Coupling$ arises from
modifications to
$\feffNewton$,
which become relevant only at small values of
$|\Coupling/\Mp|$ where the Galileon strongly
modifies the $\LCDM$ background.
This explains the nearly $\Coupling$-independent structure
in Fig.~\ref{fig:likelihood-neg}.
Throughout the $1\sigma$ region
the force law is very close to Newtonian gravity, with
an effective growth parameter $\feff \approx 0.68$.

In Figs.~\ref{fig:likelihood-lo}
and~\ref{fig:growth-factors-lo}
we plot the best-fit regions and
growth factors in a model with $\xi = 0.01$.
In this case the
constraint on $M$ strengthens slightly,
so that $M = -6 \times 10^{-3} \Mp$
is now on the boundary of the $3\sigma$ region.
Fig.~\ref{fig:growth-factors-lo}
shows that $\langle \theta\theta \rangle$, $\langle \theta\delta \rangle$
and $\langle \delta\delta \rangle$
are still closely correlated;
the effective growth rates measured from different combinations
of the two-point functions
are nearly equal.
Therefore matter concentrations
and large-scale flows are
still mostly associated with infall
into the same set of potential wells,
although these are now an 
admixture of Newtonian and Galileon contributions.
As a result of this mixing
the effective force
driving matter into the potential wells
has received a scale-dependent renormalization,
implying a departure from the $1/r^2$ force law.
This modification
drives the weakening fit with increasing $\xi$.
Clearly the effect of the second set of potential
wells is very significant, even though we do not
observe strong stochastic effects
associated with matter moving between different combinations
of them.

In Figs.~\ref{fig:likelihood-hi}
and~\ref{fig:growth-factors-hi}
we plot results for the case $\xi = 0.1$.
The model is now excluded at more than $3\sigma$.
The amplitude of the scale-dependence in the
growth rate becomes larger,
implying stronger departures from the $1/r^2$ force law,
and a small amount of decorrelation becomes visible.
Of the four cases exhibited in the plot,
this model shows the largest decorrelation
although it is still modest---of order $0.5\%$.
In Figs.~\ref{fig:likelihood-eq}
and~\ref{fig:growth-factors-eq}
we exhibit the case of equal potential wells, $\xi = 1$.
In this case the scale-dependence of the force law continues
to grow stronger,
but the decorrelation \emph{decreases} slightly.
This occurs because scale-dependence is driven entirely by the
Galileon contribution, and as it becomes dominant
the power spectra $\langle \theta\theta \rangle$, $\langle \theta\delta \rangle$
and $\langle \delta\delta \rangle$ will again become completely correlated.

\begin{figure*}
	\subcaptionbox{Irrelevant Galileon potential wells,
				   $\xi \approx 0$.
				   \label{fig:likelihood-neg}}
				  {\includegraphics[scale=0.37]{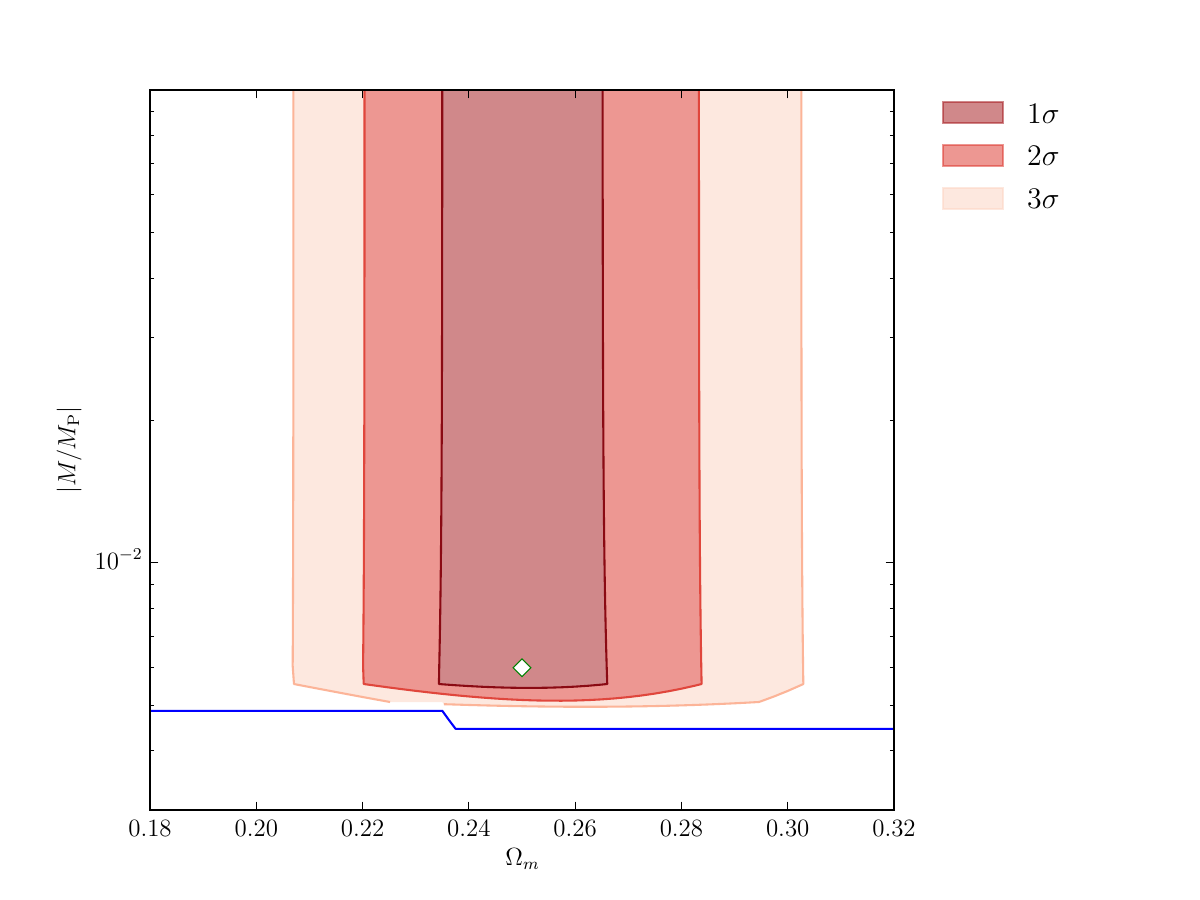}}
	\subcaptionbox{Best-fit pairwise velocity dispersion
				   $\sigma$
				   for Fig.~\ref{fig:likelihood-neg}.\label{fig:sigmav-neg}}
				  {\includegraphics[scale=0.37]{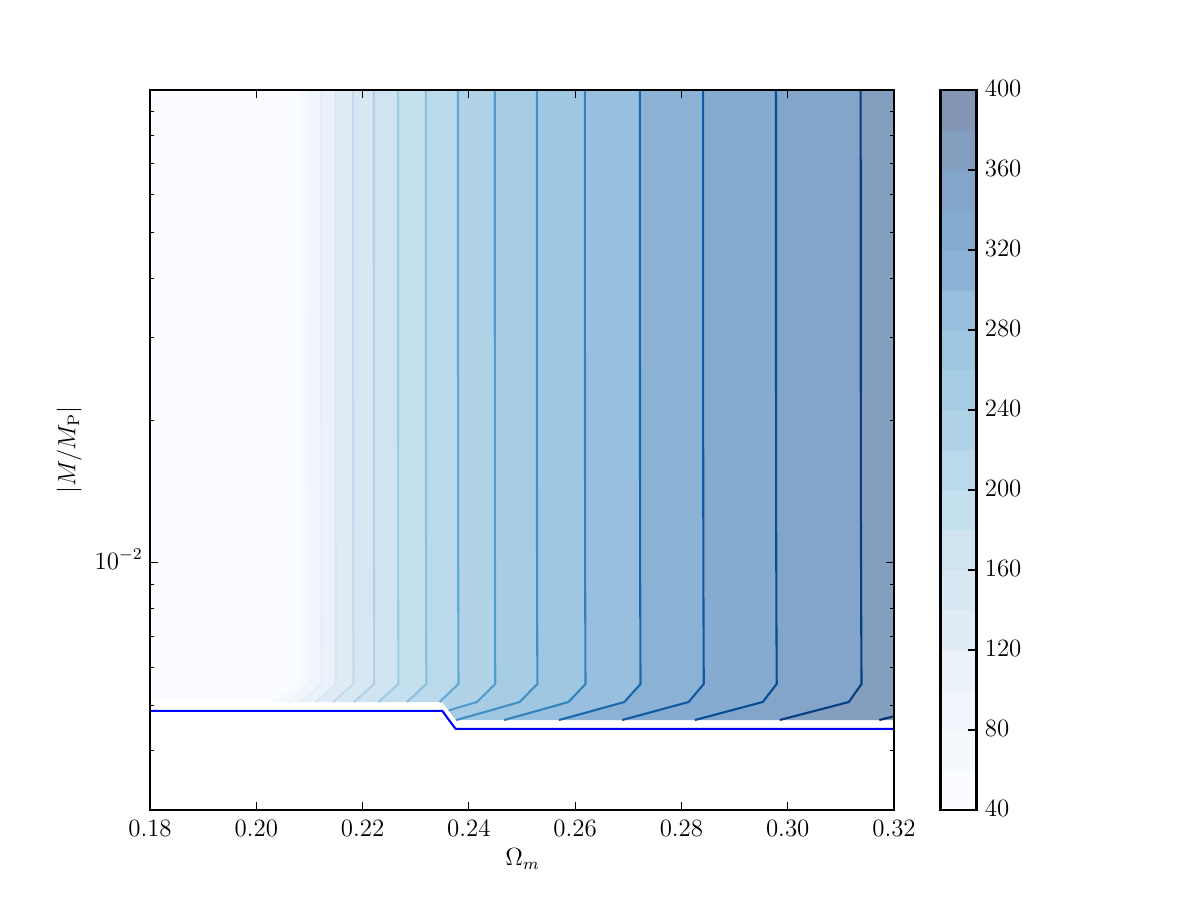}}

    \subcaptionbox{Marginal Galileon potential wells,
                   $\xi = 0.01$.
				   \label{fig:likelihood-lo}}
                  {\includegraphics[scale=0.37]{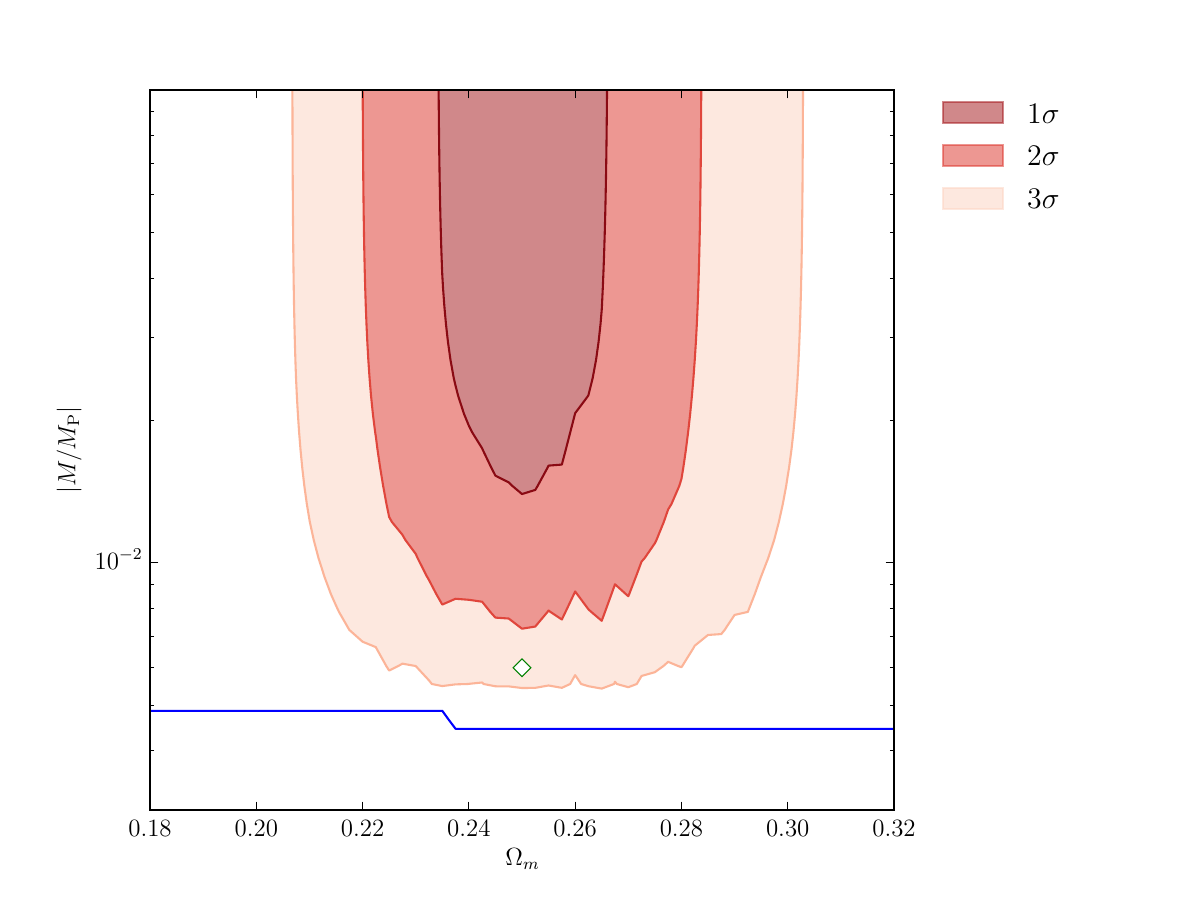}}
    \subcaptionbox{Best-fit pairwise velocity dispersion
    			   $\sigma$
                   for Fig.~\ref{fig:likelihood-lo}.\label{fig:sigmav-lo}}
                  {\includegraphics[scale=0.37]{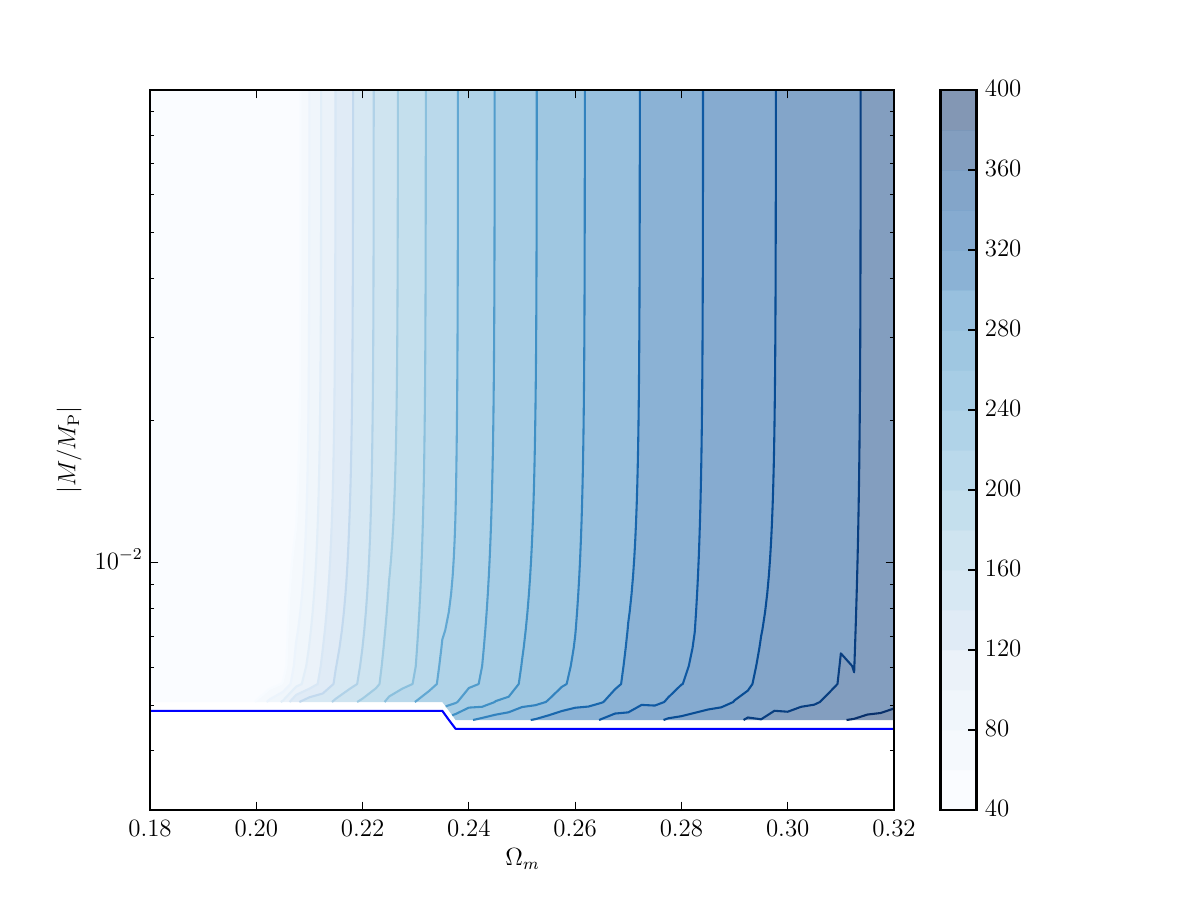}}

    \subcaptionbox{Relevant Galileon potential wells,
                   $\xi = 0.1$.
				   \label{fig:likelihood-hi}}
                  {\includegraphics[scale=0.37]{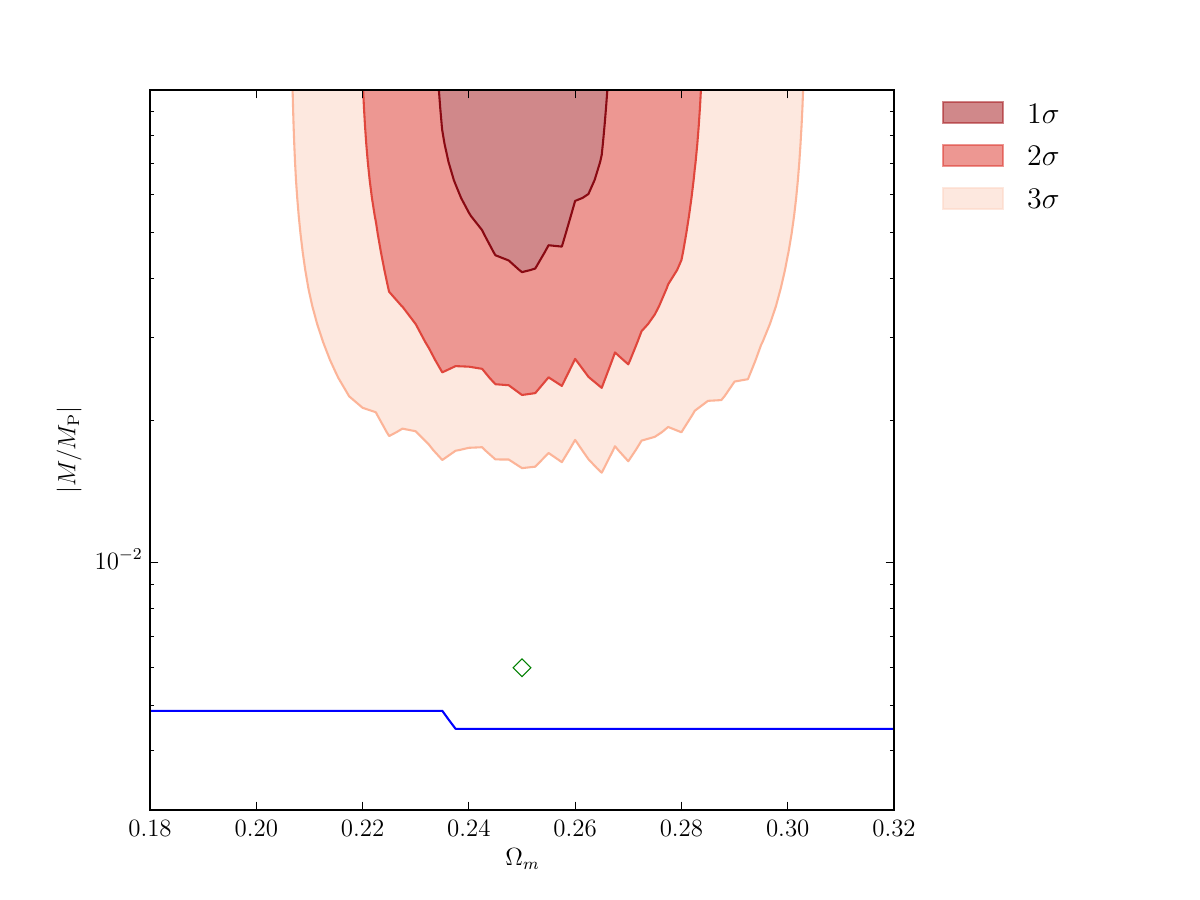}}
    \subcaptionbox{Best-fit pairwise velocity dispersion $\sigma$
                   for Fig.~\ref{fig:likelihood-hi}.\label{fig:sigmav-hi}}
                  {\includegraphics[scale=0.37]{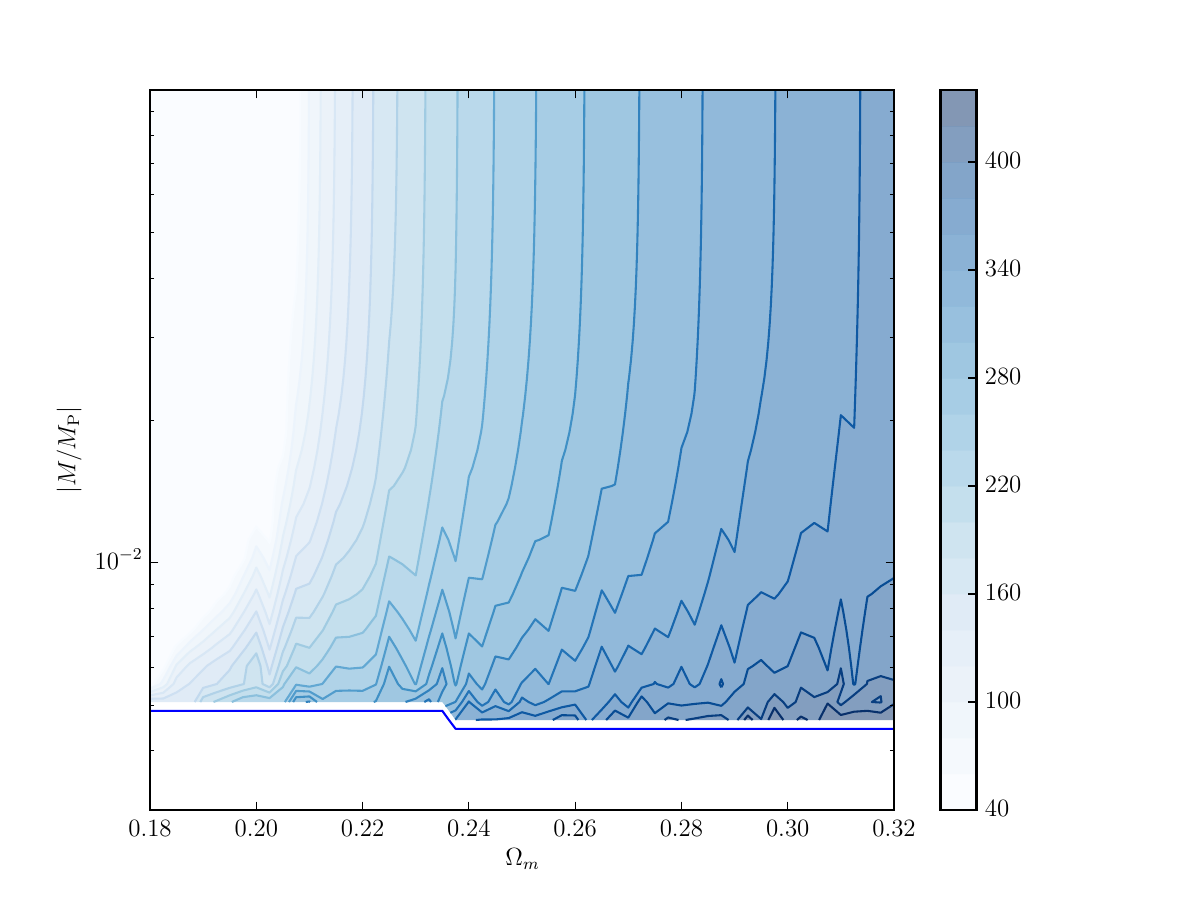}}

    \subcaptionbox{Equal Galileon potential wells,
                   $\xi = 1$.
				   \label{fig:likelihood-eq}}
                  {\includegraphics[scale=0.37]{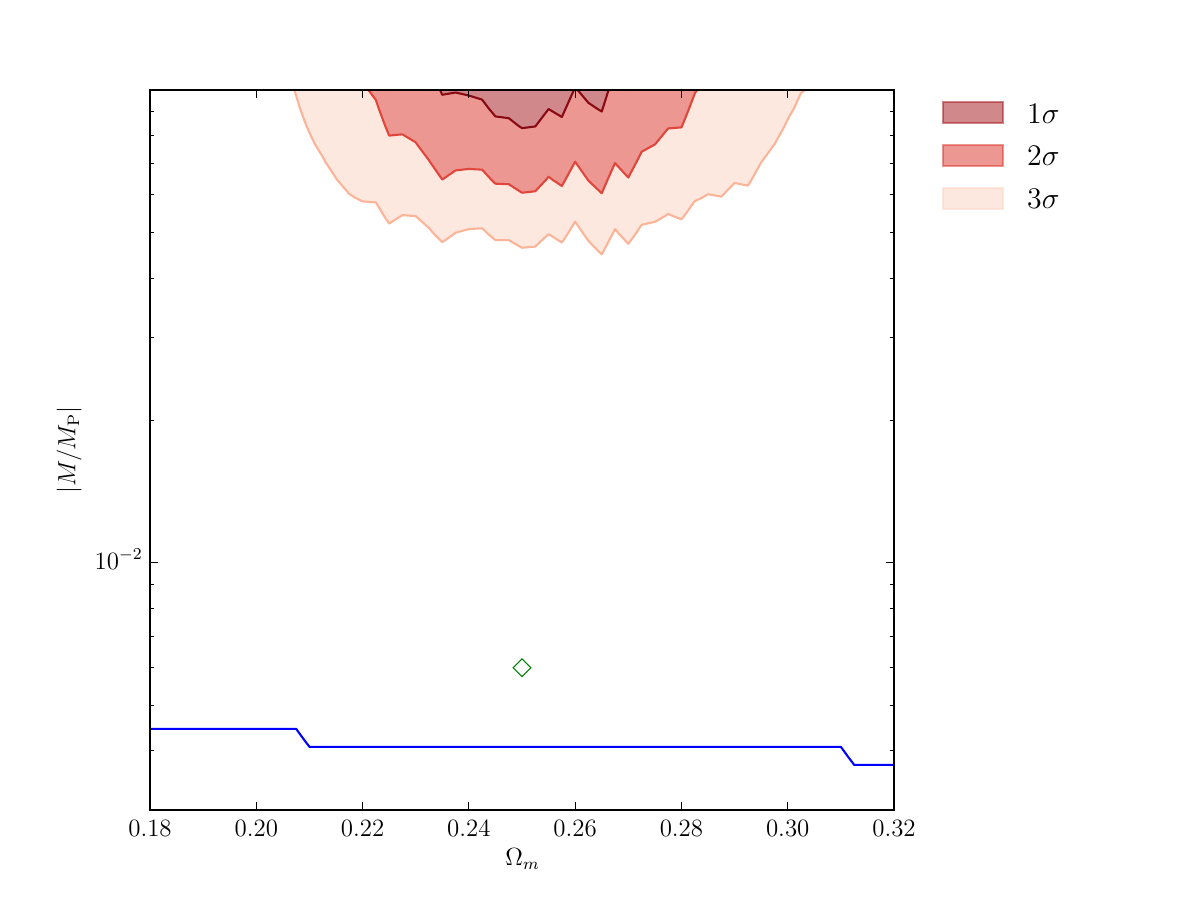}}
    \subcaptionbox{Best-fit pairwise velocity dispersion $\sigma$
                   for Fig.~\ref{fig:likelihood-eq}.\label{fig:sigmav-eq}}
                  {\includegraphics[scale=0.37]{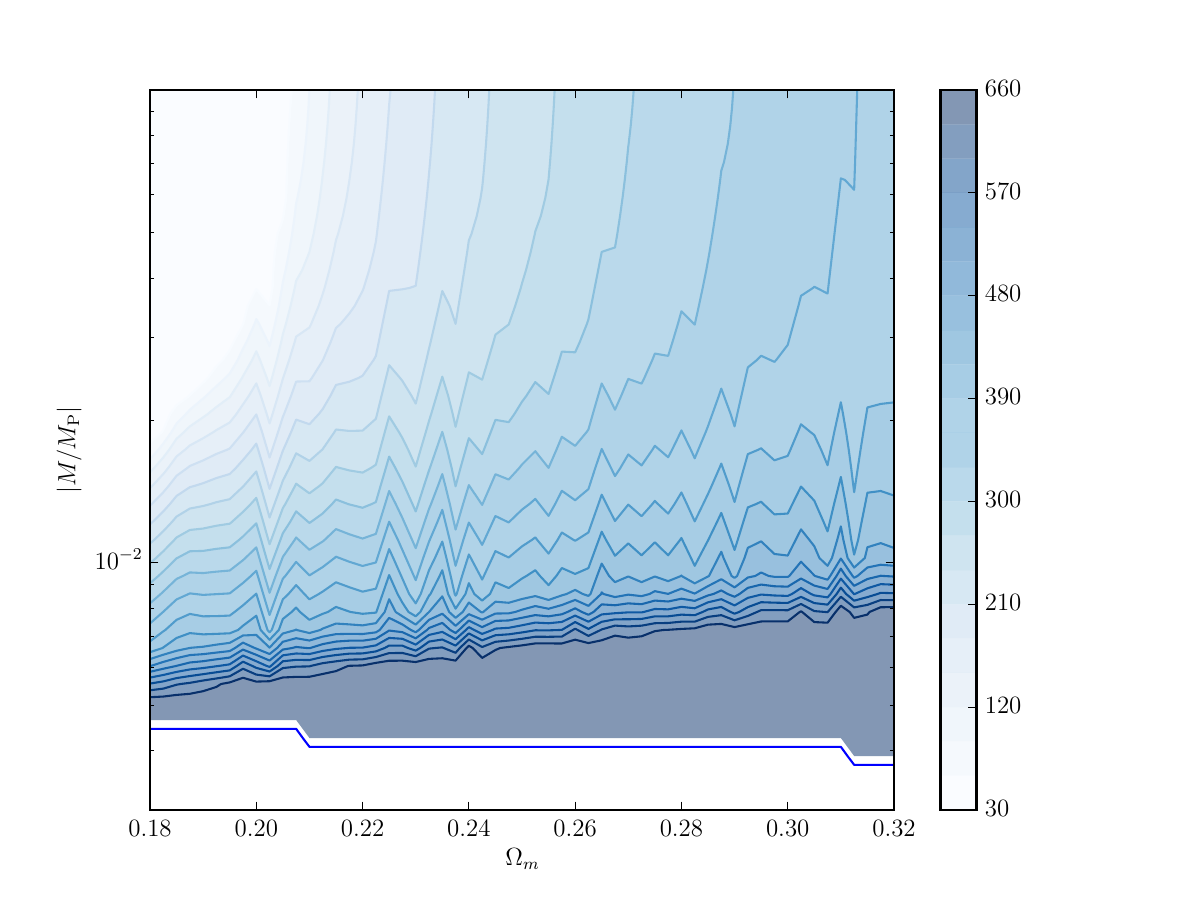}}

    \caption{Likelihoods and best-fit values for the
    pairwise velocity dispersion $\sigma$ for cubic Galileon model.
    $\xi$ measures the relative depth of the
    Newtonian and Galileon potential wells on large scales
    and was defined
    in Eq.~\eqref{eq:xi-def}.
    We exclude the region below the blue contour because
    $|\phi/\Coupling|$ becomes close to unity,
    implying that corrections to the matter coupling
    due to matter loops
    become uncontrollably large.
    The pairwise velocity dispersion is measured in
    $\mathrm{km}\,\mathrm{s}^{-1}$;
    see Eq.~\eqref{eq:kaiser}.
    Growth factors for the model indicated by the green
    diamond are given in Fig.~\ref{fig:growth-factors}.\label{fig:likelihood}}
\end{figure*}

\begin{figure*}
	\subcaptionbox{Irrelevant Galileon potential wells,
				   $\xi \approx 0$.\label{fig:growth-factors-neg}}
				  {\includegraphics[scale=0.38]{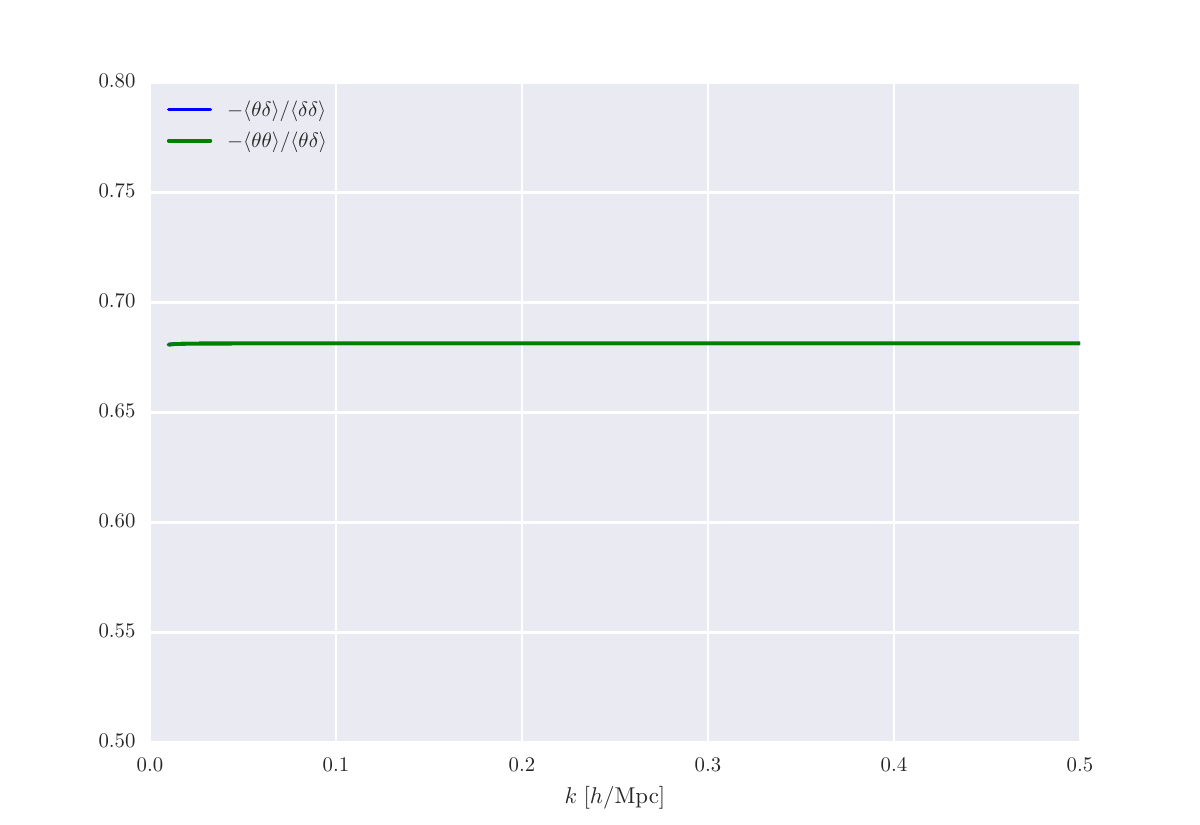}}
	\subcaptionbox{Marginal Galileon potential wells,
                   $\xi = 0.01$.\label{fig:growth-factors-lo}}
                  {\includegraphics[scale=0.38]{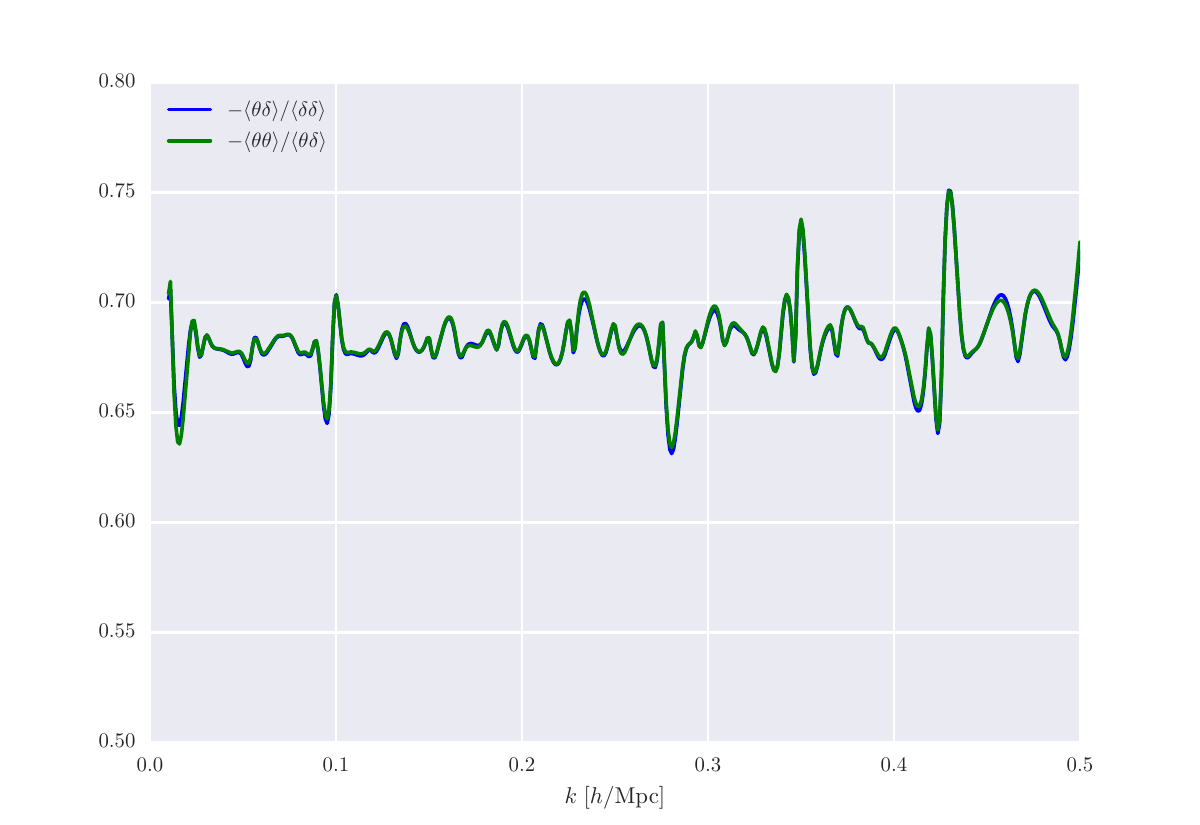}}
                  
    \subcaptionbox{Relevant Galileon potential wells,
                   $\xi = 0.1$.\label{fig:growth-factors-hi}}
                  {\includegraphics[scale=0.38]{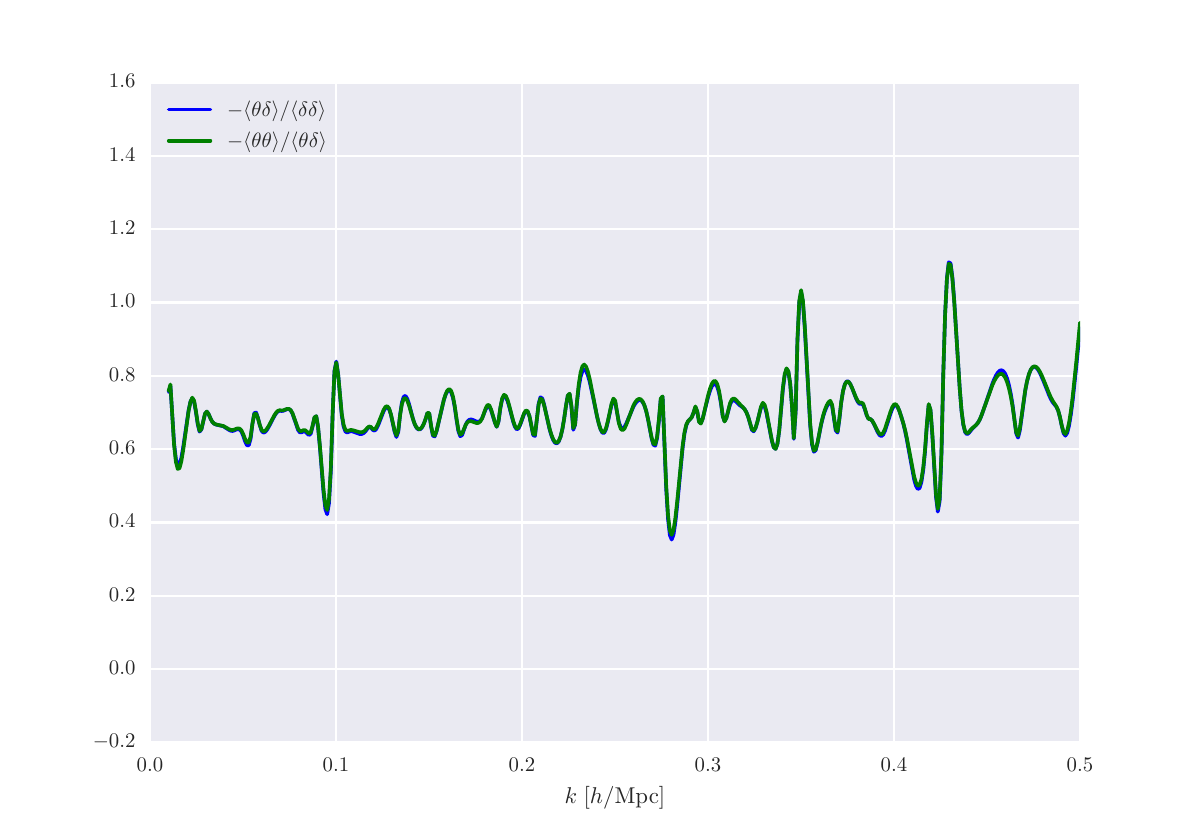}}
    \subcaptionbox{Equal Galileon potential wells,
                   $\xi = 1$.\label{fig:growth-factors-eq}}
                  {\includegraphics[scale=0.38]{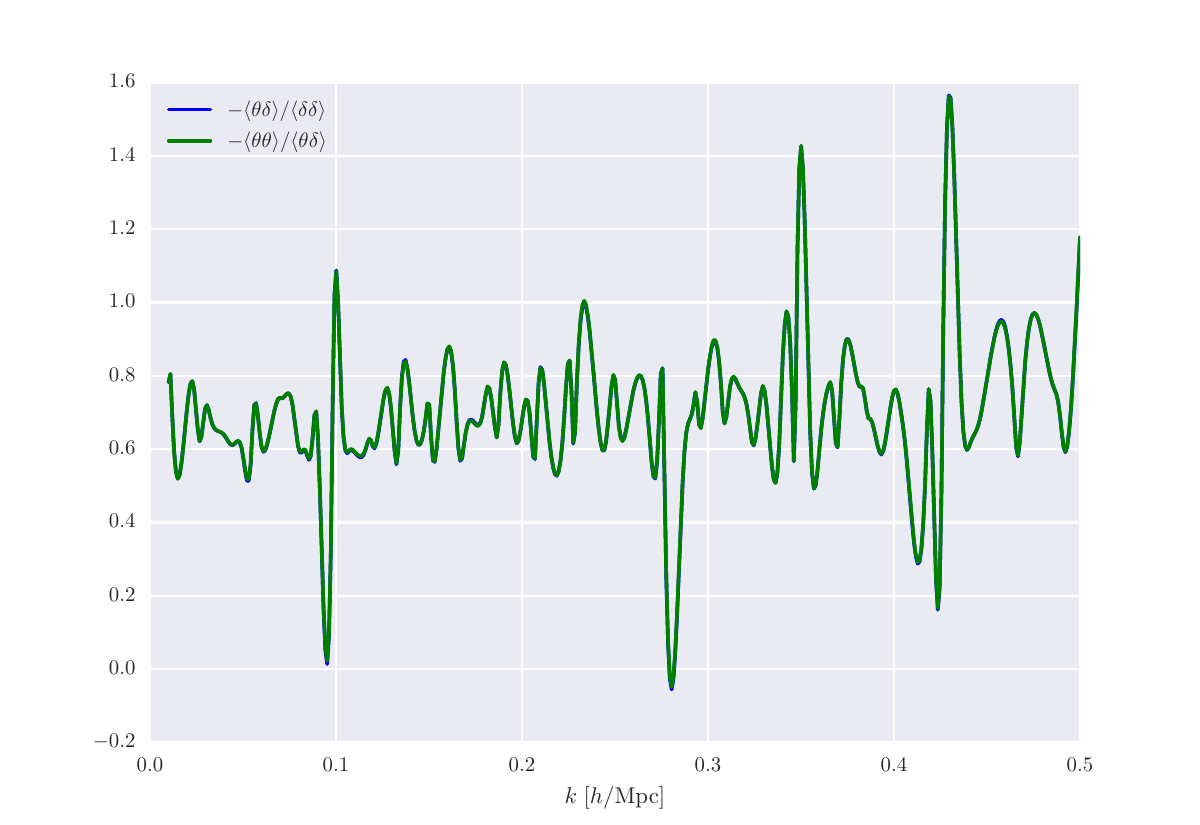}}
                  
    \caption{Effective growth rates measured
    in our numerical integrations
    at $z=0.44$ from the ratios
    $-\langle \theta\delta \rangle / \langle \delta\delta \rangle$
    and $-\langle \theta\theta \rangle / \langle \theta\delta \rangle$.
    We have taken $\Omega_m = 0.25$,
    $\SelfInteraction = 1.05 \times 10^{16} H_0$
    as in Fig.~\ref{fig:likelihood},
    and $\Coupling = -6 \times 10^{-3} \Mp$.
    This model corresponds to the diamond in Fig.~\ref{fig:likelihood}.
    The parameter $\xi$ measuring the relative depth of
    the primordial Newtonian and Galileon potential wells
    is defined in Eq.~\eqref{eq:xi-def}.
    The effective growth rates measured from
    the different 2-point functions
    are equal when $\theta$ is totally correlated with
    $\delta$, making the bias deterministic.\label{fig:growth-factors}}
\end{figure*}

\begin{figure*}
	\includegraphics[scale=0.46]{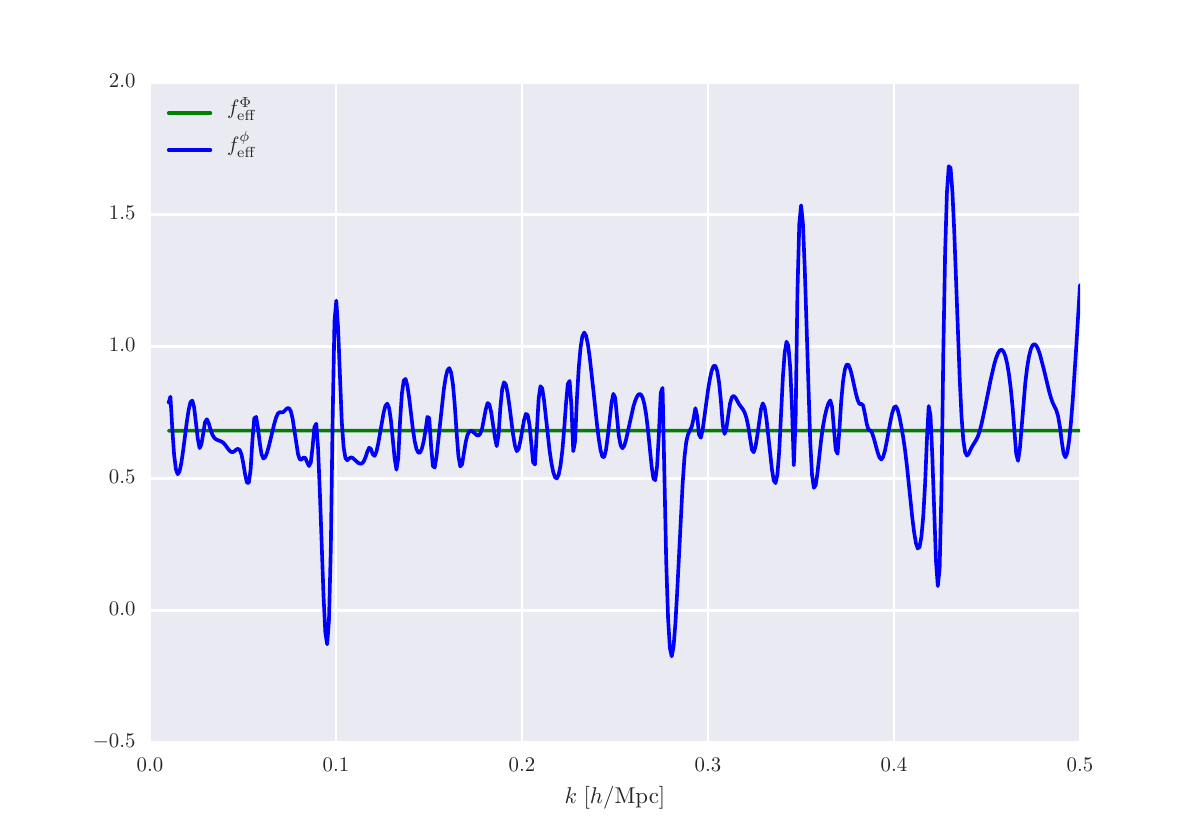}
	\caption{Growth factors $\feffNewton$ and $\feffGalileon$
	measured from our simulations at $z=0.44$
	[defined in Eqs.~\eqref{eq:feffNewton-def}--\eqref{eq:feffGalileon-def}].
	The model is the same as Fig.~\ref{fig:growth-factors}.
	The Newtonian value is $\feffNewton \approx 0.68$,
	matching the $\LCDM$ prediction $\Omega_m(z)^{0.55}$.
	\label{fig:csv}}
\end{figure*}

\para{Oscillations}
In Fig.~\ref{fig:csv}
we plot the
growth factors $\feffNewton$ and $\feffGalileon$
in this model;
the scenarios shown in Fig.~\ref{fig:growth-factors}
are weighted averages of these.
We see that $\feffNewton$ is smooth
while $\feffGalileon$ is highly oscillatory.
These oscillations are driven by rapid
fluctuations in the evolution of the Galileon field,
which would invalidate the quasi-static approximation
(see~\S\ref{sec:einstein-gravity}).

We have verified that the oscillation pattern is stable against changes in the
choice of numerical solver
(LSODA,%
	\footnote{see Ref.~\cite{hindmarsh1983odepack}; also described at \href{http://www.oecd-nea.org/tools/abstract/detail/uscd1227}{this URL}.}
VODE%
	\footnote{see Ref.~\cite{brown1989vode}; also described at \href{http://www.netlib.org/ode/vode.f}{this URL}.}
and a simple Dormand--Prince 4$^{\mathrm{th}}$/5$^{\mathrm{th}}$-order stepper),
and therefore
we believe these oscillations to be real features of the model.%
	\footnote{It is not clear that we have resolved the oscillations completely.
	We do not believe that this would change our predictions,
	precisely because they are so rapid.}
Had we imposed the quasi-static approximation we would not have resolved them
individually;
instead, we expect that the solution would predict a local average.
Observations are typically sensitive only to local averages of this kind,
because they must aggregate over a range of angles and
wavenumbers to increase signal-to-noise in each $(k, \mu)$ bin.
This washes out individual features.

For the same reason the existence of these oscillations does not
immediately falsify the model, even though current observations
are consistent with smooth power spectra.
It is possible that oscillations of this kind could be detected
by a survey such as DESI that covers much larger volumes~\cite{Aghamousa:2016zmz}.

\section{Conclusions}
\label{sec:conclusions}

\para{Summary of results}%
In this paper we have set up a formalism
to describe the relation between the
$\langle \delta \delta \rangle$,
$\langle \delta \theta \rangle$
and $\langle \theta \theta \rangle$
correlation functions in models
where $\delta$ and $\theta$ are partially or
totally decorrelated.
In such models fifth-force effects cannot
always
be absorbed into a simple
renormalization of the effective
gravitational constant,
and the $\langle \delta \delta \rangle$,
$\langle \delta \theta \rangle$
and $\langle \theta \theta \rangle$
correlation functions may no longer be related by
a fixed multiplicative normalization.
Therefore the density and velocity
fields associated with a single population
of tracers may exhibit `stochastic' bias,
depending on the combination of primordial
potentials
in which
matter is currently concentrated
and the combination into which it is moving.

We have focused on scenarios in which
the background Galileon energy density
is subdominant throughout the expansion
history of the universe.
This is the best-motivated scenario
if the Galileon arises as a vestige
of other physics which solves the
cosmological constant problem, such as degravitation
due to a graviton mass.
We have demonstrated
that, in these scenarios,
scale-dependent renormalization of the effective
force law driving cosmological mass assembly
can be constrained by
measurements of the
low-$\ell$
multipole power spectra
$P_{\ell}$
introduced by
Cole, Fisher \& Weinberg~\cite{Cole:1993kh}.
The multipole power spectra are sensitive to
some of the information
regarding density and velocity correlations
which is contained in the two-dimensional
power spectrum $P(k,\mu)$.
If enough multipoles can be measured they
also constrain the effects of stochasticity.

In this paper our constraints were obtained
using measurements of $P_0$, $P_2$ and $P_4$ from
the WiggleZ galaxy redshift survey.
In the cubic Galileon model, which we use
as a demonstration of principle,
the effects of stochasticity are not large
and constraints are largely driven by
changes to the force law due to
mass accumulating in a combination
of the primordial Newtonian and Galileon potentials.
We obtain an approximate bound
$|\Coupling| \gtrsim 5 \times 10^{-3} \Mp$
on the conformal coupling to matter in this model.

\para{Screening effects}%
The cubic Galileon model is known to have certain pathologies,
including a negative energy density
on the cosmological Vainshtein solution.
Its merit is that, on this solution, the Vainshtein radius
is independent of the model parameters
and, for clusters, roughly coincides with their physical radius.
This means that linear perturbation theory is
trustworthy down to roughly cluster scales.

In more general models the Vainshtein radius of
a cluster may be substantially larger than its physical
radius. In this case linear perturbation theory
must break down roughly at the Vainshtein radius,
which will generally depend on the physical properties
of the cluster.
For such models the validity of linear theory
becomes a stochastic statement,
depending on the realization of the density fluctuation
within a given survey volume.
It is not yet clear whether this difficulty can be
surmounted within an analytic calculation
(perhaps using the methods of extreme value statistics),
or necessarily requires recourse to $N$-body methods.

\para{Cosmological Vainshtein mechanism}%
A key role is played by the cosmological
Vainshtein solution, Eq.~\eqref{eq:cosmological-vainshtein-solution}.
This has two effects.
First, it keeps the Galileon energy density subdominant
to normal matter and radiation over most of the lifetime of
the universe. This enables the background cosmology to accurately
track $\LCDM$.
Second, it suppresses fifth-forces until late times when
the Galileon's contribution to the cosmic energy budget begins to grow.
It is the cosmological Vainshtein mechanism which is
responsible
for enabling much of the parameter space in
Figs.~\ref{fig:likelihood-neg},
\ref{fig:likelihood-lo},
\ref{fig:likelihood-hi}
and~\ref{fig:likelihood-neg}
to be compatible with observation, even though
the Galileon fifth-force is stronger than gravity
everywhere in these plots.

\para{Multiple primordial potential wells}%
Figs.~\ref{fig:likelihood-neg},
\ref{fig:likelihood-lo},
\ref{fig:likelihood-hi}
and~\ref{fig:likelihood-eq}
show substantially different behaviour,
which can be traced to the presence of
a second independent set of primordial potential
wells in Fig.~\ref{fig:likelihood-hi}.
It is mixing with this second set of potential
wells which generates the scale-dependent
renormalizations of the force law
seen in Fig.~\ref{fig:growth-factors},
even though $\delta$ and $\theta$
remain strongly correlated.

The phenomenology of these models therefore depends
on their inflationary prehistory.
If inflation seeds a distribution of Galileon
potential wells which are sufficiently deep,
they leave a detectable imprint on the observable power
spectrum.
On the other hand, if the primordial
Galileon potentials are shallow (perhaps because the
Galileon was heavy during inflation, or
because the potentials were later erased by
some form of superhorizon evolution),
their effect on the power spectrum
is negligible even if the matter
coupling is large.
Where the primordial Galileon potential wells are not
negligible
their influence must be taken into
account to obtain quantitatively correct
results, even if there is no
significant decorrelation between $\delta$ and $\theta$.

\emph{Note added}.---After completion of this paper,
a preprint was released by Jennings \& Jennings
studying stochastic effects in the relation between
$\theta$ and $\delta$
within Einstein gravity~\cite{Jennings:2015mja}.
This source of stochasticity would compete with
any effects from mixing of independent potential wells.
If it is large,
identifying stochastic effects from
fifth forces would therefore require its contribution to
be carefully subtracted.

\begin{acknowledgments}
CB is supported by a Royal Society University Research Fellowship.
DP was supported by an Australian Research Council Future Fellowship [grant number FT130101086].
DS acknowledges
support from
the Science and Technology Facilities Council
[grant numbers ST/I000976/1 and ST/L000652/1]
and the Leverhulme Trust.
Numerical computations were performed on the Sciama High Performance Compute (HPC) cluster which is supported by 
the ICG, SEPNet and the University of Portsmouth.

We would like to thank Chris Blake for providing the WiggleZ
multipole power spectrum dataset used in this analysis.
CB and DS thank the Astrophysics group at the University of Queensland
for their hospitality during the early stages of this collaboration.

\para{Data availability}
No new data were collected for this paper.
Transfer functions and power spectra used in constructing
Fig.~\ref{fig:2dps},
and for the model comparison in
Fig.~\ref{fig:likelihood},
can be obtained as SQLite databases
from a permanent deposit at
\href{http://zenodo.org}{zenodo.org}
using DOI
\href{http://doi.org/10.5281/zenodo.343840}{10.5281/zenodo.343840}~\cite{dataset}.
The WiggleZ multipole power spectrum dataset
used to construct the likelihood
was provided by the WiggleZ team;
for current contact details,
see the survey homepage at
\href{http://wigglez.swin.edu.au/site/index.html}{http://wigglez.swin.edu.au/site/index.html}.

\end{acknowledgments}

\bibliography{refs}

\end{document}